\documentclass[preprint,authoryear,5p]{elsarticle}  

\newif\ifpreprint
\preprinttrue 

\usepackage{graphicx}
\usepackage{amsmath}
\usepackage{amssymb}    
\usepackage{epsfig}
\usepackage[latin1]{inputenc}
\usepackage{hyperref,url}
\usepackage{placeins}
\usepackage{xcolor}

\ifpreprint
\else
	\usepackage{lineno}  
	\linenumbers
\fi

\usepackage[T1]{fontenc}
\usepackage{ae,aecompl,url}
\usepackage{multicol}

\title{Accretion of Ornamental Equatorial Ridges  on Pan, Atlas and Daphnis}

\author[a1]{Alice C. Quillen\corref{cor1}}
\ead{alice.quillen@rochester.edu}
\author[a1]{Fatima Zaidouni}
\ead{fzaidoun@u.rochester.edu}
\author[a1,a2]{Miki Nakajima}
\ead{mnakajima@rochester.edu}
\author[a1]{Esteban Wright}
\ead{ewrig15@ur.rochester.edu}
\address[a1]{Department of Physics and Astronomy, University of Rochester, Rochester, NY 14627 USA}
\address[a2]{Department of Earth and Environmental Sciences, University of Rochester, Rochester, NY 14627, USA}
\cortext[cor1]{Corresponding author}

\hypersetup{draft}

\begin{document}

\begin{abstract}
We explore scenarios for the accretion of ornamental ridges on Saturn's moons Pan, Atlas, and Daphnis from material in Saturn's rings. Accretion of complex shaped ridges from ring material should be possible when the torque from accreted material does not exceed the tidal torque from Saturn that ordinarily maintains tidal lock. This gives a limit on the maximum accretion rate and the minimum duration for equatorial ridge growth. We explore the longitude distribution of ridges accreted from ring material, initially in circular orbits, onto a moon that is on a circular, inclined or eccentric orbit.   Sloped and lobed ridges can be accreted if the moon tidally realigns during accretion due to its change in shape or because the disk edge surface density profile allows ring material originating at different initial semi-major axes to impact the moon at different locations on its equatorial ridge. We find that accretion from an asymmetric gap might account for a depression on Atlas's equatorial ridge.  Accretion from an asymmetric gap at orbital eccentricity  similar to the Hill eccentricity, might allow accretion of multiple lobes, as seen on Pan. Two possibly connected scenarios are promising for growth of ornamental equatorial ridges.  The moon migrates through the ring, narrowing its gap and facilitating accretion. The moon's orbital eccentricity could increase due to orbital resonance with another moon, pushing it into its gap edges and facilitating accretion.
\end{abstract}


\maketitle

\section{Introduction}

\begin{figure}  
\centering
\includegraphics[width=3.5in]{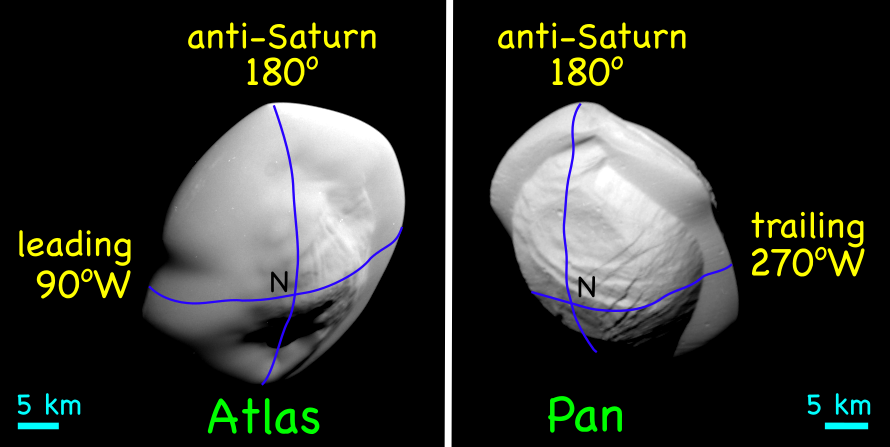} 
\includegraphics[width=3.5in]{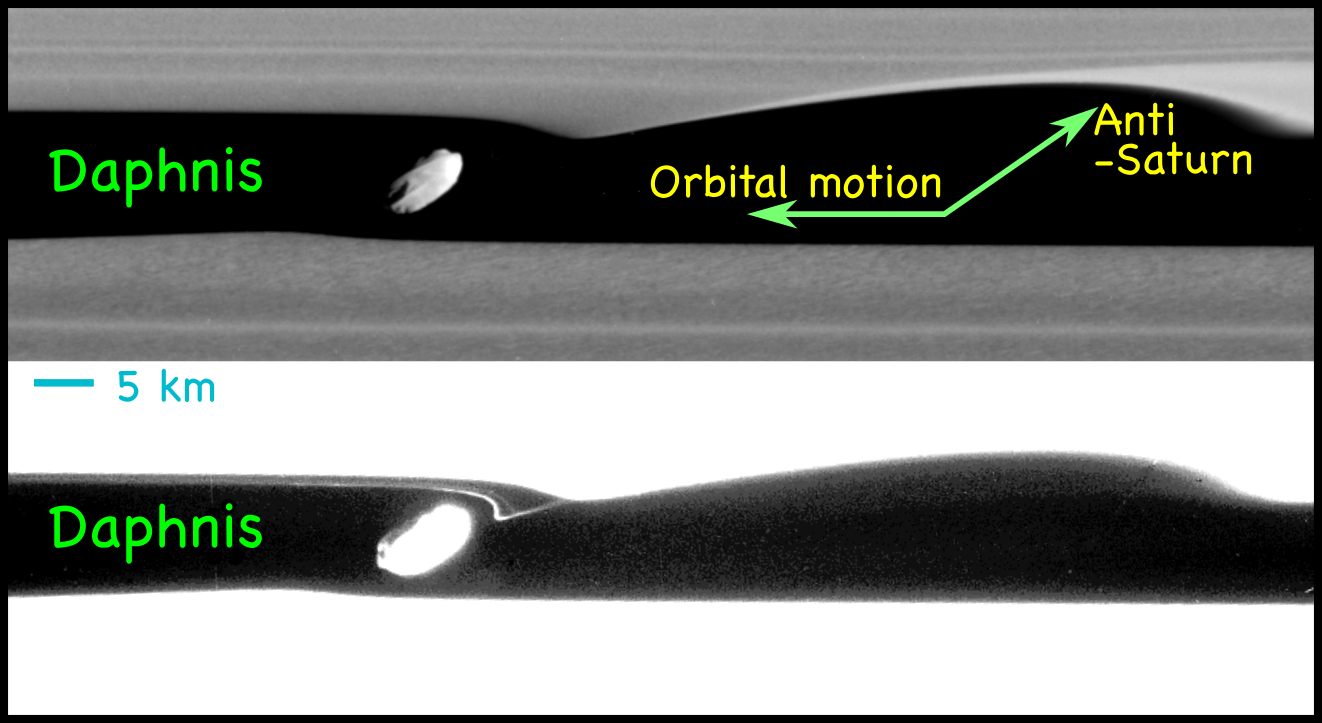} 
\caption{
Grayscale images of the moons Pan, Atlas, and Daphnis
obtained with ISS (the Imaging Science Subsystem) 
during the late Cassini  mission flybys.  
The image numbers are N1867604669 for Pan,  
N1870699120 for Atlas, and N1863267232 for Daphnis. 
Cyan bars show 5 km distances.
We labelled longitudes with blue lines on Pan and Atlas 
using the wire frame figures by \citet{thomas20} as a guide.  These are  shown in their Figures
7 and 8,  and derived from their shape models.
Convention is that $0^\circ$ longitude on the moon faces Saturn and 
$90^\circ$ W faces the direction of the moon's orbital motion.  
The north poles, labelled 'N' on Pan and Atlas are located where the two blue lines cross.
The bottom two panels show the same image of Daphnis,
but with different contrasts
The bottom image better 
shows a faint wisp interior to the ring edge of material that was perturbed by Daphnis.
These moons
nearly fill their Roche region.  Body axis ratios are listed in Table \ref{tab:moon}. 
\label{fig:moons}}
\end{figure}

Toward the end of the  Cassini mission, the spacecraft made a series of close approaches
to Saturn's rings that included flybys of moons Pan, Daphnis, Pandora, Atlas, and Epimetheus.
The images taken from these close approaches increasingly revealed that the small inner moons
of Saturn have irregular and intricate shapes \citep{thomas10,thomas13,buratti19,thomas20}.  
We focus here on three moons with intricate equatorial ridges, dubbed `accretion ornaments'
\citep{charnoz07}.
Pan has a sharp edged `ravioli'-shaped equatorial ridge,  Atlas has a thin, lens-shaped equatorial ridge with a depression on one side.  Daphnis is elongated and two or three two narrow ridges cross its body (see Figure \ref{fig:moons}).  

The surfaces of Pan's and Atlas' equatorial ridges are smoother than the more rounded central or core components of each moon which show grooves or fractures \citep{buratti19,thomas20}.  
This may also be true on Daphnis but the spatial resolution of the Daphnis imagery is poorer than that of Pan and Atlas, so it is more difficult to determine whether Daphnis' ridges are comprised of material similar to its underlying core.
The central core components exhibit more impact craters than do the equatorial ridges on Pan and Atlas, which nevertheless display a few sub-kilometer impact craters \citep{buratti19}.
At high and low latitudes, Pan's core appears bare of loose material.  In contrast, Atlas's core seems 
covered by about ten meters of loose material which has been called `icy regolith' \citep{buratti19}.
Pan's equatorial ridge encircles its equator and resembles a polygon \citep{buratti19,thomas20}.

The shapes of Pan and Atlas are not close to gravitational equilibrium figures or Roche ellipsoids \citep{porco07,thomas20}.  
These moons are not spinning fast enough to form a diamond or top shape like some asteroids (e.g., asteroid 1999 KW4; \citealt{ostro06}).
Tidal stress due to Saturn can deform a moon, 
altering its surface morphology \citep{buratti19}.  
However, tidal stress 
would elongate a synchronously rotating body in the radial direction (along the moon-Saturn line) rather than create a thin equatorial ridge.  Neither centrifugal nor tidal forces alone can account for all aspects of the equatorial ridge morphology (e.g., see discussion by \citealt{charnoz07,leleu18,thomas20}). 
Using SPH simulations, \citet{leleu18} showed that the spectrum of shapes of Saturn's small moons is a natural outcome of slow merging collisions. 
Head-on mergers result in flattened objects with  equatorial ridges, like Atlas and Pan, whereas with more oblique impact angles, mergers result in elongated bodies like Prometheus  \citep{leleu18}.  However mergers are unlikely to account for the differing core and ridge surface morphologies.
The equatorial ridges most likely formed through slow accretion of ring particles \citep{charnoz07}
and both slow impacts and subsequent accretion could have taken place.

Cross sections of Pan's and Atlas' equatorial ridges are shown in Figure 9 by \citet{thomas20}.
Pan's  ridge has a maximum width, measured in the direction
perpendicular to its equator, 
of about 6 km.  However  PIA images N1867606181 and N1867606643 of Pan 
(see Figure 7 by \citealt{thomas20})
show that the ridge's outer edge is quite sharp and less than 1 km thick.
Most of Altas' ridge (between 180 and 360$^\circ$ W longitude, as shown
in the map in Figure 10 by \citealt{thomas20}) is less than 8 km thick. 
Daphnis'  ridges, as delineated in Figure 3B by \citet{buratti19},
have width less than 1/2 km thick.
Ridges with sharp edges would be consistent with formation via accretion from fine low inclination ring material \citep{charnoz07}.  The widths of these equatorial ridges
are consistent with the low moon orbital inclinations and deposition from ring
particles with height distribution in the rings near these moons (see Table 4 
and associated discussion by \citealt{thomas20}).  
Images of Pan in Figure \ref{fig:moons} and in  the supplements by \cite{thomas20} 
show that its equatorial ridge exhibits thickness variations
but suggest that the equatorial ridge on Pan is not warped. 
There are only a few high resolution images of Pan taken when the spacecraft was nearby
so it is difficult to determine whether thickness variations are similar or different on each  lobe.

Recent measurements of Pan's, Daphnis', and Atlas'  
orbital elements, mass, density, orbital period, and body semi-major axes \citep{porco07,jacobson08,buratti19}
are compiled in Table \ref{tab:moon}.  We also list the volume of the equatorial ridge divided by total moon volume, $f_v$, estimated by \citet{buratti19} and \citet{thomas20}.
 
All three moons lie within Saturn's main rings.  
Daphnis orbits Saturn in the Keeler Gap which has a half-width
of 21 km and lies within the A ring. 
Pan, also in the A-ring,   is found in the Encke Gap  \citep{showalter91} which has  a half-width of 161 km.
Atlas lies outside the A-ring in the Roche Division.
The gravitational field of Saturn itself dominates the orbital motions of these moons and a  precessing Keplerian ellipse gives an adequate model for their orbits  \citep{jacobson08}. 
Atlas has a significant orbital eccentricity and inclination compared to the lower values for Pan or Daphnis, and this is due to gravitational perturbations from Prometheus and Pandora  \citep{spitale06,cooper15}. 



Prior work on accretion onto moons in ring systems  includes a numerical study of accretion onto embedded moonlets where the moonlet's Roche lobe remains filled with accreted material
(\citealt{yasui14}; also see Figure 4 by \citealt{porco07}).  
Here a `moonlet' is a body with Hill radius less than a kilometer that is 
not massive enough to open a gap in the ring \citep{bromley13}.
Pan, Atlas and Daphnis are large enough to open gaps in the ring, 
so we refer to them as moons rather than moonlets.  
\citet{charnoz07} proposed that accreting ring material would primarily 
flow through Pan's L1 or L2 Lagrange points onto Pan where it would
impact the moon equator preferentially on Pan's Saturn-facing and anti-Saturn  sides. 
\citet{charnoz07} proposed that the divots in
Atlas's equatorial ridge are caused by  
its orbital eccentricity.  In their numerical simulations of an accreting
moon on eccentric orbit, more accreting material 
hit the surface on the moon's leading
side than on the trailing side.  

Pan's remarkable 5 or 6 pointed ravioli-shaped equatorial ridge has remained a curiosity.  
Likewise the multiple ridges crossing Daphnis lack a dynamical explanation.
In this study we explore processes that might cause accreting ring material
to pile up at particular longitudes on a moon's equator.   
We  follow the model for equatorial ridge formation proposed by \citet{charnoz07}, 
where the equatorial ridge grows via slow accretion from  ambient ring material.
We do not pursue alternate possible scenarios for equatorial ridge growth, such as 
those involving   a minor merger \citep{leleu18} or an  accretion disk that falls back onto
the moon equator 
(e.g., relevant for Iapetus; \citealt{ip06,stickle18}). 

In section \ref{sec:torque} we estimate torques on the three moons.
Accretion scenarios are sensitive to orbital eccentricity, so
we discuss orbital eccentricity damping via driving
of spiral density waves in section \ref{sec:edamp}. 
In section \ref{sec:corb} we examine the impact angle of  material accreting
from a cold ring onto a moon that is in a circular orbit.  We also explore  
the morphology of an equatorial
ridge that is accreted while tidally locked and from material that impacts
the moon surface at a single impact angle.  We discuss accretion 
from the edge of a gap in the ring, with gap edge  described with a surface density profile.  
In section \ref{sec:eorb} we examine the distribution
of impact angles for material accreting from a ring onto a moon that is in an eccentric orbit.
Scenarios for equatorial ridge accretion are discussed in section \ref{sec:scen}.
A summary and discussion is in section \ref{sec:sum}.
\ref{ap:mom} describes additional details used in our illustrative toy models.

\section{Tidal and accretion torques}
\label{sec:torque}

We consider a moon that is accreting from a stream of material that impacts
its surface.   If the accretion stream impacts the moon tangentially, (it approaches
on a non-radial trajectory with respect to the moon's center of mass),
then the accreting material exerts a torque on the moon that 
can perturb the moon's spin rate and orientation. 
Saturn's regular moons are in spin-synchronous
or tidally locked states due to the tidal torque exerted on them by Saturn.
We estimate the torque on an initially spinning moon and 
the time it takes to reach tidal lock via internal dissipation.  We compare the tidal torque to 
the torque that could arise from accretion.
If the torque from accreting material is high enough, it would pull the moon out of 
the spin-synchronous state.  
 
\subsection{Tidal torque}
\label{sec:spindown}

The secular part of the semi-diurnal ($l = 2$) term in the Fourier expansion of the perturbing potential from point mass $M_*$,  gives a tidally induced torque on an extended body of mass $M$ 
\begin{equation} 
T_2 = \frac{3}{2} \frac{GM_*^2}{a}  \left( \frac{R}{a} \right)^5 k_2 \sin \epsilon_2, \label{eqn:Tdown}
\end{equation} 
where  $k_2 \sin\epsilon_2$ is known as the quality function, describing deformation and dissipation in $M$ 
 (e.g., \citealt{kaula64,goldreich63,efroimsky13}).
The quality function is often approximated as $k_2/Q$ where $Q$ is a dissipation factor for mass $M$ that is 
commonly called the `Quality factor'.  The  Love number $k_2$  characterizes tidal deformation of $M$.
Here $a$ is the orbital semi-major axis of $M$ about $M_*$,   $R$ is the radius
of mass $M$, assumed spherical, and $G$ is the gravitational constant.
The torque causes the spin $\omega$ of $M$ to vary.  The torque 
can be used to estimate a time $t_{\rm despin}$ for a spherical body to spin up or down from 
an initial spin rate of $\omega_0$ to the 
tidally locked state.  Assuming an initial spin rate near  
centrifugal break-up $\omega_0 = \sqrt{GM/R^3}$, 
\begin{equation}
 t_{\rm despin} \sim  \frac{\omega_0}{\dot \omega}
\sim  \frac{2P}{15 \pi} \left(\frac{M}{M_*} \right)^\frac{3}{2}
\left( \frac{a}{R} \right)^\frac{9}{2} \frac{Q}{k_2} , \label{eqn:t_despin}
\end{equation}
where $P$ is the initial orbital period. 
We use this time to characterize the rate that a moon's spin varies
when it is not in a spin-synchronous (tidally locked) state or within a spin-orbit resonance.
For estimates of this time scale,  
we adopt  Love number $k_2  = 0.038 e_g/\mu_{\rm shear}$ (consistent with the derivation 
by \citealt{frouard16} for a homogeneous viscoelastic sphere) 
where gravitational binding energy $e_g \equiv GM^2/R^4$
and $\mu_{\rm shear}$ is the shear modulus of mass $M$'s interior.
We adopt $\mu_{\rm shear} Q = 10^{11}$ Pa typical of ice or rubble \citep{pravec14}.
This value is consistent with $\mu_{\rm shear} = 10^9 $ Pa and $Q=100$.
The tidal spin-down times for the three moons are computed from values listed
in Table \ref{tab:moon} for the three moons and are listed
in Table \ref{tab:computed}.
The spin-down times, $\sim 100,000$ yr,  are short compared to
the age of the Solar system, but are long compared
to the orbital periods of Saturn's regular moons. 

\subsection{Torque from accretion}
\label{sec:torque_acc}

If a moon is accreting mass at a rate $\dot M$, approximately what is the torque on the moon?
The L1 or L2 Lagrange equilibrium points are about 1 Hill radius $R_H = a (\mu/3)^\frac{1}{3}$
away from the moon center of mass.  Here $\mu \equiv M/M_*$ is the ratio of moon to planet mass.
To estimate the torque, we consider the spin angular momentum 
imparted to the moon by the impact of an unperturbed test particle in an initially coplanar, circular orbit about the planet. 
A test particle that is in a circular orbit 
at semi-major axis of $a'  = a + \Delta a$ 
with separation $\Delta  a = R_H$ has 
 mean motion 
 $n' = n + \frac{dn}{da} R_H$.
The difference in mean motion between
test particle and moon  $\delta n = \frac{dn}{da} R_H = \frac{3}{2} n \frac{R_H}{a}$
(and we are ignoring the sign of our estimate).  
The tangential velocity component (computed using the
origin at the moon's center of mass) of the test particle is
$v_\theta \sim a \delta n \sim \frac{3}{2} v_H$ where $v_H \equiv n R_H$ is the Hill velocity. 
If this particle hits and sticks to the moon, the 
change in spin angular momentum of the moon per unit accreted mass would be  
\begin{equation}
\delta l \sim R_H v_\theta \sim  \frac{3}{2} v_H R_H  \sim  \frac{3}{2} n R_H^2 .
\end{equation}
The torque on the moon due to accreting material is approximately
\begin{align}
T_{\rm acc} \sim \frac{3}{2}  \dot M n R_H^2 . \label{eqn:Tacc}
\end{align}
Below we integrate orbits that impact
a moon (sections \ref{sec:corb} and \ref{sec:eorb}) to improve upon this torque estimate.  

The torque from accreting material can vary the moon's spin, possibly even pulling it out of 
the spin synchronous state if the accretion rate is high enough.
What accretion rate 
is large enough for the torque from accreting material to exceed the spin-down tidal torque from Saturn on the moon? 

The ratio of the torque from accreting material (equation \ref{eqn:Tacc}) to that caused by Saturn's tide (if not tidally locked; equation \ref{eqn:Tdown}) is
\begin{align}
 \frac{T_{\rm acc}}{T_2} &\sim \frac{\dot M}{M n}
 \left(\frac{M}{M_*}\right)^\frac{5}{3} 
 \left(\frac{a}{R}\right)^5 
 \left(\frac{Q}{k_2}\right)  3^{-\frac{2}{3}}, \label{eqn:Tacc_ratio}
\end{align}
where we have replaced the quality function with $k_2/Q$.
We recognize  an accretion time scale  $t_{\rm acc} \equiv M/\dot M$.
In  Table \ref{tab:computed} we include values for the torque ratio
$\frac{T_{\rm acc}}{T_2} \frac{Mn}{\dot M}$
computed for the three moons using dissipation factor
$Q$ and Love number $k_2$ computed with $\mu_{\rm shear} Q = 10^{11}$ Pa as previously
described in section \ref{sec:spindown}.
 
When the torque from accreting material exceeds the tidal torque from Saturn, 
the moon  would be pulled out of a tidally locked
 spin synchronous state or a spin-orbit resonance.
What mass accretion rate is sufficient to allow this to happen?
We set $T_{\rm acc} = T_2$, balancing accretion with tidal torque, and solve equation \ref{eqn:Tacc_ratio} 
for a critical accretion $\dot M_{cr}$
\begin{align}
\dot M_{cr} = M n  \left(\frac{M_*}{M}\right)^\frac{5}{3}
 \left(\frac{R}{a}\right)^5 \left(\frac{k_2}{Q}\right) 3^\frac{2}{3}. \label{eqn:Mcrit}
\end{align}
Critical values of the accretion rate are listed in Table \ref{tab:computed}
for the three moons 
and are approximately $\dot M_{cr} \sim 10^{10}$ kg/yr.
We also include  in Table \ref{tab:computed} estimates for an accretion  timescale 
at this rate 
of the equatorial ridges   $M f_v/\dot M_{cr}$ (where $f_v$ is the fraction
of volume in the ridge).
 
Accretion rates above $\dot M_{cr}$ would be large enough to pull the moon
out of the spin synchronous state and in that case we would expect smooth equatorial ridges that
lack divots or polygon shapes.  As the equatorial ridges in Pan and Atlas have non-axisymmetric
structure,
we would infer that the accretion rates during equatorial ridge formation
were lower than the critical ones.
We estimate that the time over which accretion took place was longer 
than the time scale $M f_v/\dot M_{cr}$ which is of order $10^5$ years for these moons (see Table \ref{tab:computed}). 
This implies that the duration for accretion of these equatorial ridges was probably
longer than about $10^5$ years.  

We compare the critical accretion rate to viscous accretion rate through the A ring.
Using a ring viscosity of $\nu = 64\ {\rm cm}^2/{\rm s}$ in the A-ring near Pan \citep{tajeddine17b},
and a ring mass surface density of $\Sigma_{\rm ring} \approx 40\ {\rm g/cm}^{2}$ 
\citep{tiscareno07}, 
the viscous accretion  rate through the ring  at Pan's orbital semi-major axis 
$\dot M_{\rm cross} \sim 3 \pi \Sigma_{\rm ring} \nu \sim 10^{9}$ kg/yr.
Our estimate for the critical value $\dot M_{cr}$ is above this value, so if accretion
onto Pan is fed purely by viscous accretion in a ring edge we could expect accretion 
to take place below the critical accretion rate $\dot M_{cr}$.  

A larger rate of mass could cross the moon's orbit if the moon migrates.  \citet{bromley13} estimate that Pan and Daphnis have masses in the range that allows rapid migration to occur, at about $\frac{d a_{\rm mig} }{dt}\sim 20$ km/yr.   
Rapid or type III migration occurs when orbit crossing material exchanges angular momentum with a planet or moon near the ends of horseshoe orbits \citep{masset03}.  
A condition for rapid or runaway migration, via a type III mechanism involving
co-orbital material \citep{masset03}, is proximity to dense ring material with a low dispersion of eccentricity.
At a migration rate  of $\frac{d a_{\rm mig} }{dt}\sim 20$ km/yr, a mass rate of 
$\dot M_{\rm cross} = 2 \pi \Sigma_{\rm ring} a \frac{d a_{\rm mig}}{dt} \sim 7 \times 10^{15}$  kg/yr 
crosses the moon's orbit as it drifts through the ring, vastly exceeding the critical rate, $\dot M_{cr}$. 
Most of this material would cross the moon's orbit from one side to the other side  
rather than be accreted onto the moon.
Such a high migration rate cannot be maintained in the same direction for long
(or even long enough to accrete the equatorial ridges),  otherwise the moons would have left the ring.
The two mass rates for $\dot M_{\rm cross} $ (estimated via viscous evolution and via rapid migration) bracket the possible accretion rates during formation of these equatorial ridges.
 
In summary, we estimate the torque on a moon
due to angular momentum in accreting material and we compare this torque to the 
tidal torque exerted by Saturn.  This accretion torque lets us estimate a critical accretion rate that
would be large enough to pull the moon out of a tidally locked state or a spin-orbit
resonance.  If the moon is not tidally locked (or in a spin-orbit resonance), then we would expect
accretion to be evenly distributed around the equator, giving a ridge that lacks divots
or radial extensions. 
As both Pan and Atlas's ridges have non-axisymmetric structure, we infer
that the accretion rates of their ridges primarily occurred at rates below $M_{cr}$ 
which we estimate is of order $10^{10}$ kg/yr.
The accretion rate during much of the formation of Pan's, Atlas' and Daphnis' equatorial ridges
was likely below $10^{10}$ kg/yr and took place  in a nearly tidally locked setting and with duration longer than about $10^5$ years.
We return to possible connections between disk evolution, moon migration and equatorial ridge accretion in section \ref{sec:scen}.

\section{Eccentricity damping via driving spiral density waves}
\label{sec:edamp}

The orbital eccentricity of a moon affects its mode of accretion from ring material
\citep{charnoz07}.
The eccentricity of a tidally locked body in eccentric orbit (following \citealt{peale78,yoder82})
is damped.
With tidal torque alone, it would take longer than about $10^{12}$ years 
for the orbital eccentricity of these moons to decay.  However, the torque from driving spiral density waves into the  ring exceeds the tidal torque. 

\citet{hahn08} estimated the time for Pan's eccentricity to damp  
due to spiral density waves driven by Pan into the ring material interior and exterior to Pan.
With the current Encke gap width (with a half-width $\Delta a/R_H \approx 8.45$),
\citet{hahn08} finds an eccentricity damping time $\tau_e = \dot e/e \sim 2000$ years.  
N-body simulations of a Pan- or Daphnis-sized moon in proximity to a particulate disk show
rapid eccentricity damping in about 100 years \citep{bromley13}, supporting the short
time estimated by \citet{hahn08}.
The eccentricity damping time is short, suggesting that
Pan's orbital eccentricity should be extremely small.
Pan's current non-zero eccentricity can be attributed to a nearby 16:15 inner mean motion resonance with Prometheus \citep{spitale06,hahn08}.   
Because of the short estimated eccentricity damping time, it would be difficult
to maintain Pan's eccentricity over $10^5$ years (so that it remains tidally locked during ridge accretion) 
unless Pan's eccentricity were resonantly
excited by another moon.  

Pan is inside the orbit of Prometheus.  If Pan's orbital semi-major
axis were only 31 km lower, it would be strongly perturbed by  
 the 16:15 resonance with Prometheus. 
Pan's orbital migration is currently thought to be prevented by resonant
excitations and orbital stirring of the ring material by other moons \citep{bromley13}. 
However, it is likely that Pan and other moons have migrated in the past \citep{bromley13}. 
Pan's eccentricity could have been excited in the past, possibly for
long periods of time, if it remained in an orbital resonance with another moon.

Atlas, since it currently resides  outside the A ring  (43 Hill radii away from
the edge), primarily drives spiral
density waves distant from itself and so should have a significantly longer
eccentricity damping timescale. However, during an epoch when it accreted ring material, it too could have required eccentricity
excitation to maintain its eccentricity.  

Damping times for Pan's and Daphnis' orbital inclination via excitation of density waves
are estimated to be similar in duration as those for eccentricity damping  \citep{hahn08,weiss09}.

\begin{figure*} 
\centering
\ifpreprint
\centering
\includegraphics[width=3.5in, trim={0mm 0mm 0mm 0mm}]{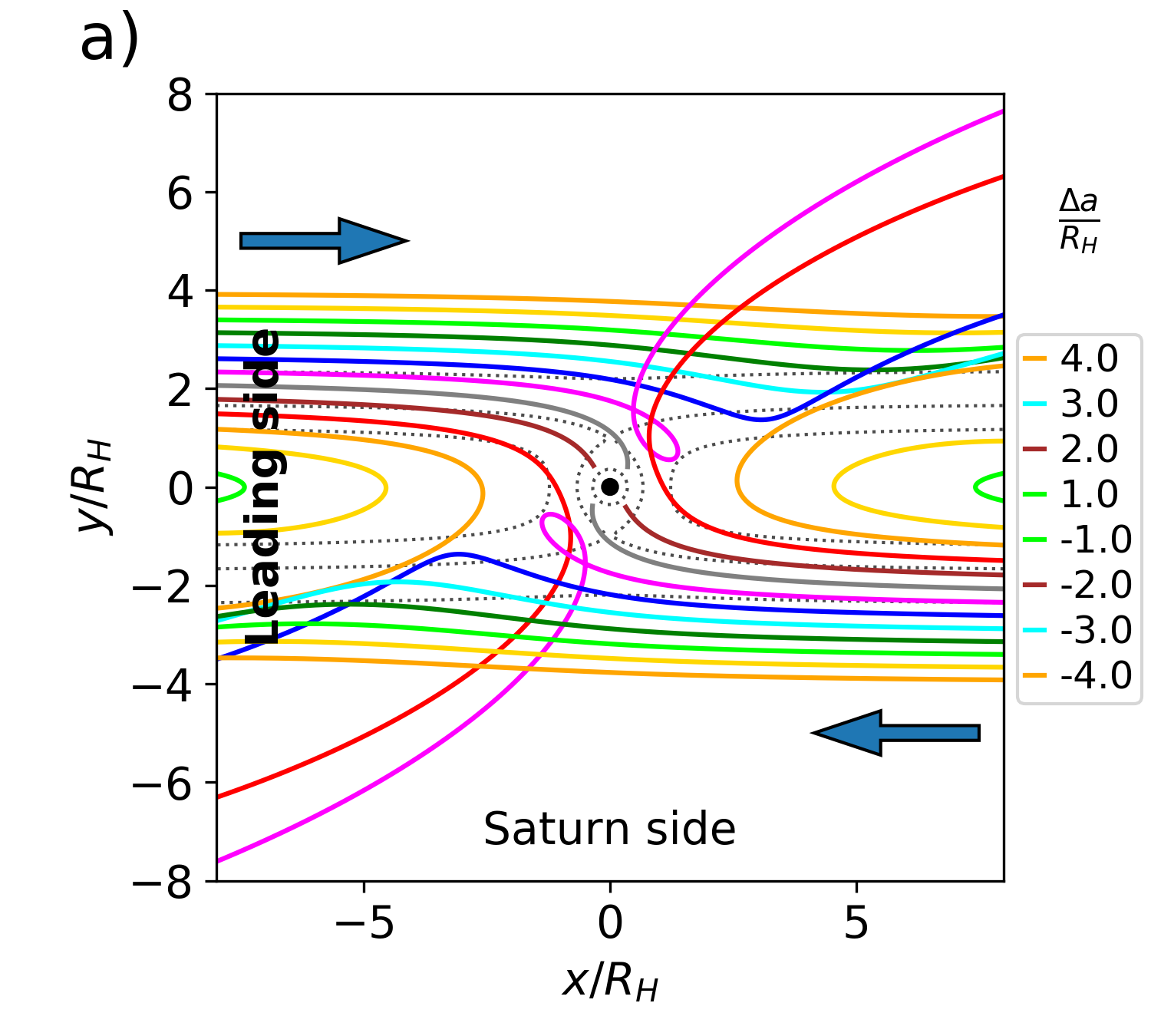} 
\includegraphics[width=3.5in, trim={0mm 0mm 0mm 0mm}]{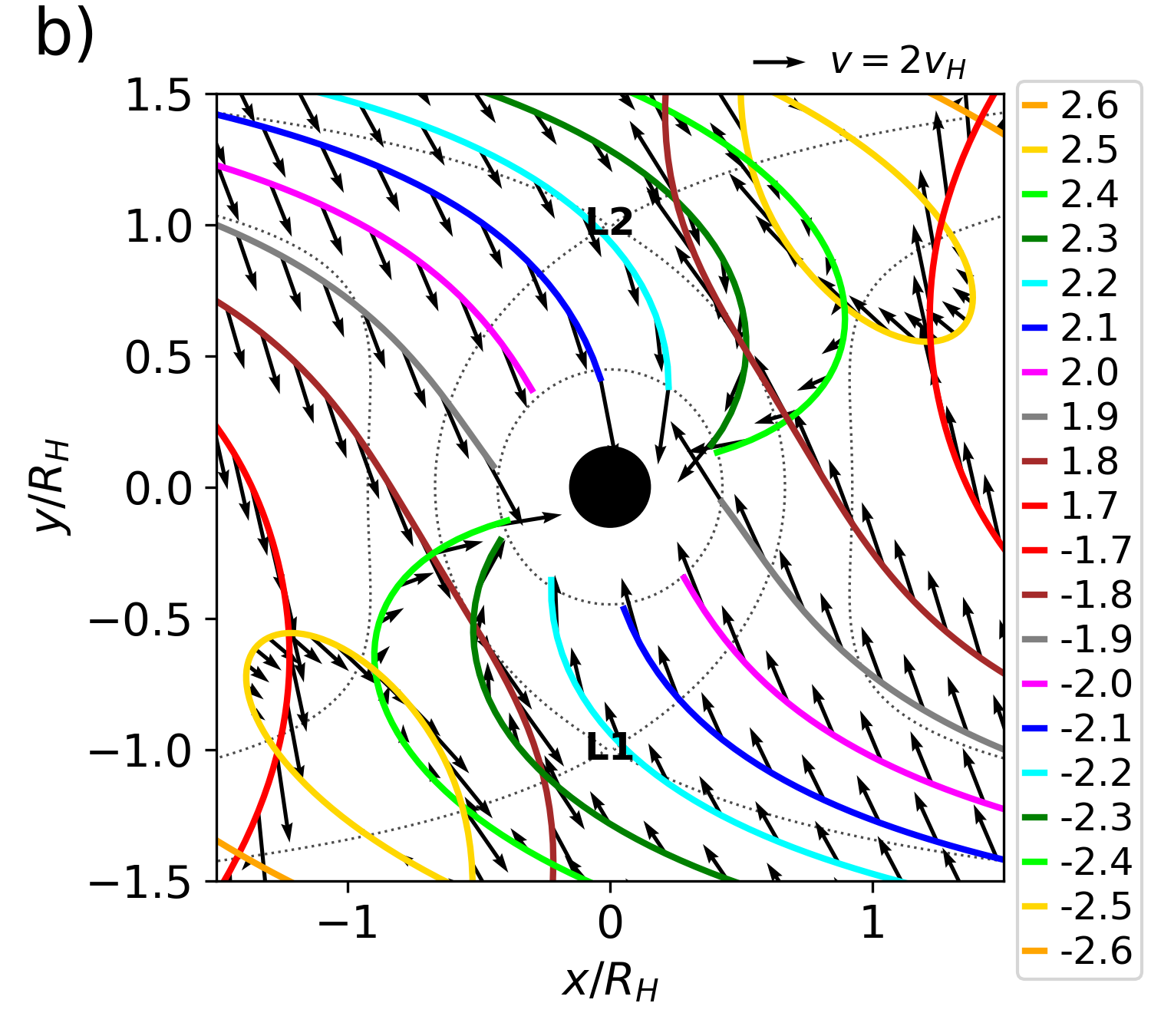}
\else
\centering
\begin{tabular}{cc}
\includegraphics[width=3.0in, trim={5mm 0mm 0mm 0mm}]{orb_mu11.png} & 
\includegraphics[width=3.0in, trim={5mm 0mm 0mm 0mm}]{quiver_mu11.png}
\end{tabular}
\fi 
\caption{Test particle orbits in the frame rotating with a moon that is in a 
circular orbit about a planet.
The moon's center of mass is shown with a black dot in the center of  each plot.
The $x$ and $y$ axes are  in units of the Hill radius and the moon is centered at the origin.
The planet is located on the negative $y$ axis. The leading side of the moon is on the left.
Test particles were begun at zero eccentricity
and inclination 
and at different orbital semi-major axes in orbit about  the central planet.
The orbits were computed with satellite-to-planet mass ratio $\mu = 10^{-11}$, however,
the morphology of these orbits in units of $R_H$ is not  dependent on the mass ratio $\mu$ as long
as this ratio is low. 
Each test particle orbit is shown with a different color line and is labelled in the key 
by its initial semi-major axis in units of $R_H$ from that of the moon's orbit. 
Grey dotted contours show levels of the effective potential, $V_{\rm eff}$ (see equation \ref{eqn:Veff}).  The saddle points of the effective potential  
are the L1 and L2 Lagrange points, with L1 on the bottom side which faces the planet.
a) We  show the test particle orbits.
Thick blue arrows show the initial direction of motion of the orbits in the rotating frame.
b) Similar to a) except showing the region closer to the moon and velocity vectors.
Velocity vectors are differences between the instantaneous velocity of the orbiting particle 
and the moon.  A vector on the upper right shows a velocity with magnitude twice the Hill velocity
$v_H$.  
Test particle impact sites onto the moon depend on 
their initial semi-major axis and how much of the Hill sphere
is filled by the moon.   We find that test particles on initially planar circular orbits
can impact the moon at almost any location on
the moon's equator.  
\label{fig:corb}}
\end{figure*}

\begin{figure} 
\centering\includegraphics[width=3.5in, trim={0mm 0mm 0mm 0mm},clip]{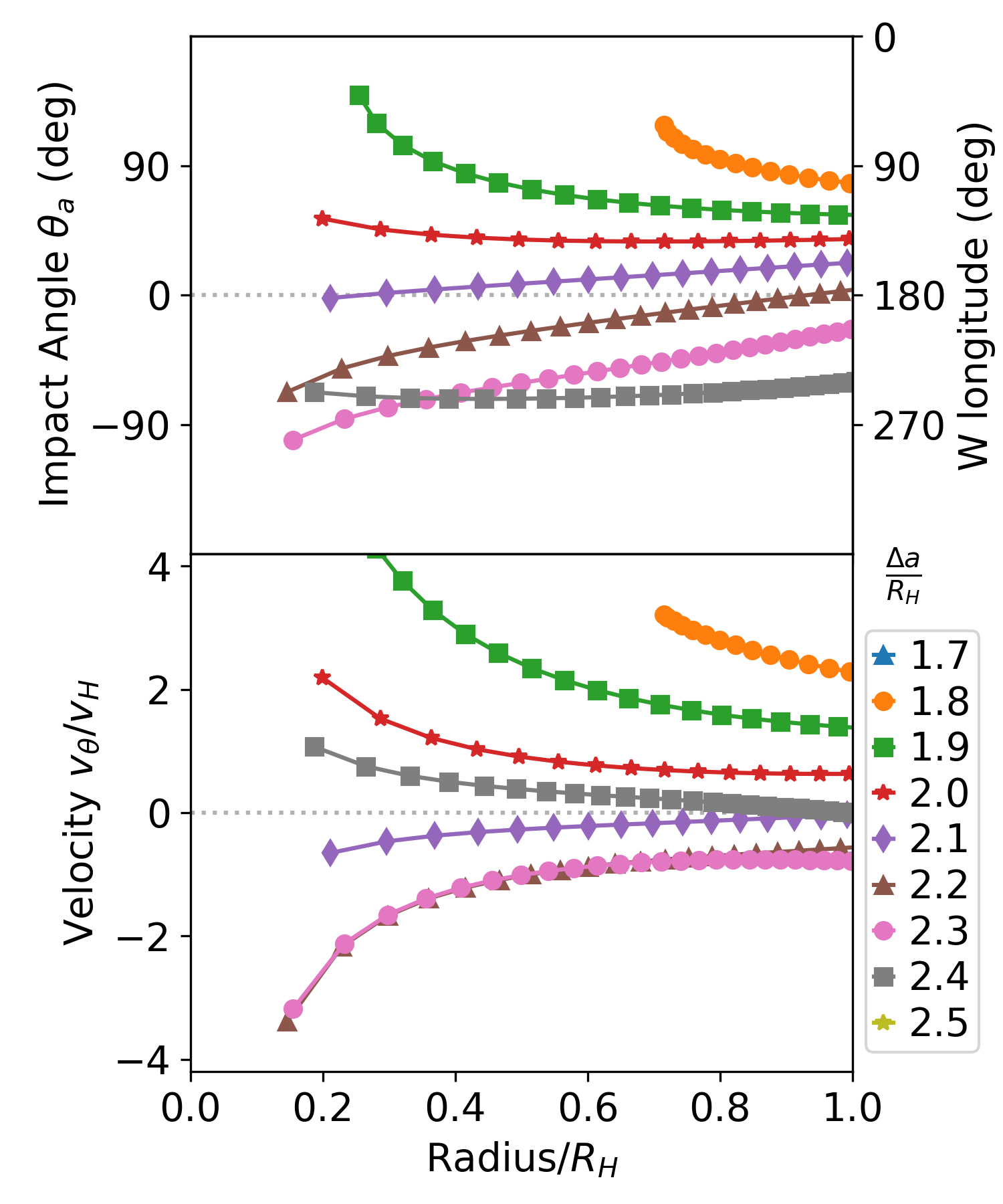} 
\caption{Angle and tangential velocity of test particle trajectories.
The top panel shows impact angle, measured from the moon/planet line ($+y$ axis), 
as a function of radius from the moon center for test particle orbits 
shown in Figure \ref{fig:corb}b.
This angle determines the impact longitude (on the moon's equator) of accreting material
(see Figure \ref{fig:pan_sil} for an illustration of this angle).
The bottom panel shows the tangential velocity component for the same orbits as a function of radius
from the moon and is in units of the Hill velocity $v_H$.    The
tangential velocity component times the radius gives the spin angular
momentum of accreting material. Accreted material with a positive value of $v_\theta$ would spin up the moon.
The key shows the different offsets $\Delta a$ between initial test particle orbit semi-major axis  and
that of the moon in units of the Hill radius $R_H$.   We  only show particles originating in orbits
exterior to the moon's orbit.  Particles originated interior to the  moon's orbit
would have impact angles rotated by $180^\circ$ from that at the same
$|\Delta a|$ but originating exterior to the moon's orbit.
Test particles that did not come within 
a Hill radius of the moon on their first approach are not plotted.  \label{fig:ang}}
\end{figure}

\begin{figure} 
\centering\includegraphics[width=3.0in, trim={0mm 0mm 0mm 0mm},clip]{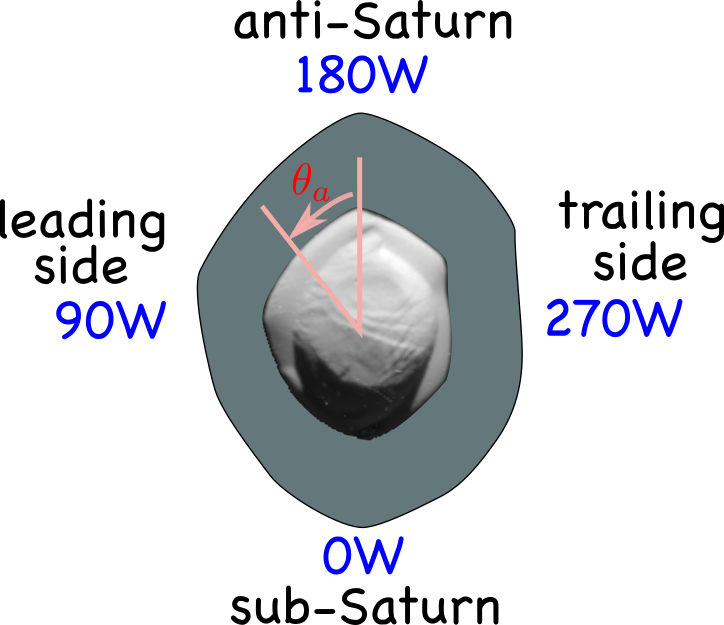} 
\caption{Our impact or accretion angle $\theta_a$  with a grey silhouette by \citet{thomas20} of Pan's
equatorial ridge as seen from above its North pole and looking downward onto Pan.  
We also mark conventionally used planetocentric West longitudes.
Inside the silhouette is the image N1867602962 of Pan which was observed at 
a ring elevation of $71^\circ$.  
\label{fig:pan_sil}
}
\end{figure}

\section{Test particles that impact a moon in a circular orbit}
\label{sec:corb}

In this section we discuss the trajectories of particles that impact a moon that is in a circular orbit.
This is a fiducial case, but this case may also be relevant for accretion while 
a moon is migrating through a disk.  In that setting
the moon may reside within a narrow gap.  A stream of fresh ring material can continually approach the moon (e.g., see Figure 4 by \citealt{bromley13} showing an N-body simulation
of a Daphnis-sized moon undergoing rapid migration through a particle disk).

We refer to massless particles as test particles. 
We integrate test particle orbits using the N-body code \texttt{rebound} \citep{rebound} 
and with the 15-order adaptive step size integrator known as `IAS15' \citep{rein15}.
The orbits are integrated in the plane of a two-body system with test particles
begun in circular orbits but at different initial orbital semi-major axes.
One massive particle is a moon and it orbits the other massive particle, the planet,
in a circular orbit.  
The orbits were integrated with a moon-to-planet mass ratio  $\mu = 10^{-11}$
similar to that of Pan.  However, 
the morphology of these orbits in units of the Hill radius
is not strongly dependent on the mass ratio, as long as $\mu$ is small.
We verified this insensitivity by making the same plots for mass ratio $\mu=10^{-7}$.
The initial conditions for the test particles were computed from orbital
elements using Jacobi or center of mass coordinates.  The test particles
were initially thousands of Hill radii away from the moon.

Test particle orbits in the frame rotating with the moon are shown in both panels of Figure \ref{fig:corb}.
Orbits were begun at orbital semi-major axes both inside and outside the moon's orbit.
The axes show distances in units of the moon's Hill radius, $R_H = a (\mu/3)^\frac{1}{3}$, with origin at the location of the moon's center.  Here $a$ is the moon's
orbital semi-major axis.  Figure \ref{fig:corb}a shows a larger region
than Figure \ref{fig:corb}b.  In Figure \ref{fig:corb}b  we  show test particle velocities subtracted by the moon's orbital velocity as vectors.
Both positions and velocity vectors have been rotated so that the moon-planet line lies
along the $y$-axis.  The planet is found below, on the negative $y$-axis and the leading
side of the moon is to the left.
The central position of the moon is shown with a black dot on both panels in Figure \ref{fig:corb}.
With grey dotted lines we show effective potential contours 
\begin{equation}
V_{\rm eff} = - \frac{r_{cm}^2}{2a^2} - \frac{(1-\mu)}{r_p/a}  - \frac{\mu}{r_m/a}, \label{eqn:Veff}
\end{equation}
where $r_{cm}$ is the radius from the center of mass of the two-body system,
$r_p$ is the radius from the planet, and $r_m$ is the radius from the moon.
The effective gravitational potential for the restricted 3-body
problem is computed in the rotating frame, with angular rotation rate that is 
in units of the mean motion $n$ of the moon.
The saddle points of the effective potential are the L1 and L2 Lagrange points
at about 1 Hill radius from the moon, with L1 on the side of the moon facing the planet.

Each test particle orbit in Figure \ref{fig:corb} is plotted with a different
color line and the lines are labelled in the keys by the difference between
their initial orbital semi-major axis and that of the moon.   This offset, $\Delta a$, 
is labelled in units of Hill radius, $R_H$.
The test particle orbit shapes in Figure \ref{fig:corb} are consistent with the orbits of 
ring particles initially in circular orbits that are integrated using Hill's 
equations \citep{dermott81a,hedman13}. 

\citet{charnoz07} proposed that accreting ring material would primarily flow through the L1
and L2 Lagrange points, at a distance of about 1 Hill radius
away from the moon center and above and below the moon in Figure \ref{fig:corb}.    
This implies that accreting material from ring material inside Pan's orbit 
 would primarily impact
Pan on the Saturn side after going through the L1 Lagrange point, 
and the opposite would be true for accreting material entering from outside Pan's orbit 
and entering through the L2 Lagrange point.
The numerical simulations by \citet{charnoz07} confirmed this 
expectation as they showed that impacts on Pan and Atlas occur preferentially near sub- and anti-Saturn points on the moon's equator.
In contrast, our Figure \ref{fig:corb} shows 
the impact locations could be almost anywhere 
on the moon's equator and the impact point depends on the initial orbital radius of the test particle.
If the accreting material comes from the edge of a cold ring, with low velocity eccentricity dispersion,  
then the impact longitude on the moon is dependent upon the distance of the ring edge from the moon's orbit.  
 
Figure \ref{fig:corb} shows that only particles with initial conditions
restricted to a range of initial semi-major axes, or $\Delta a$, would impact the moon.
Test particles with small initial offset $|\Delta a|/R_H \lesssim 1.7$ have orbits
initially similar to horseshoe orbits.  They don't
impact the moon on their first close approach to the moon.  If the orbit integrations were carried out
 for longer periods of time, some might impact the moon \citep{dermott81a}.    Our
test particles are initially in circular orbits to mimic the role of collisions in the ring.  In high-opacity rings, collisions swiftly reduce particle eccentricity to a level lower 
than $10^{-8}$ 
\citep{rein10}.

Particles in initially circular orbits which can impact the moon upon first approach are restricted
to a small range in initial orbital semi-major axis, $1.7 \lesssim |\Delta a|/R_H \lesssim 2.5$,
confirming similar numerical results by \citet{karjalainen07} (see their Figure 1).  
Within this range, test particles initially external to the moon and at larger semi-major axis impact
the moon on the trailing side, whereas those at lower $\Delta a$    
impact the moon on the leading side, as expected \citep{charnoz07}.  Orbit trajectories 
initially within the moon's orbit are similar to those initially outside the orbit 
after rotation by $180^\circ$.

In Figure \ref{fig:ang}'s top panel, we record test particle trajectory impact angles 
and in Figure \ref{fig:ang}'s bottom panel we show 
tangential velocity components as a function of radius from the moon center in 
units of Hill radius.
The longitude of an impact depends on the radius of the moon surface or how
much of the Hill sphere is filled by the moon itself.
The impact angle is counter-clockwise from the positive $y$ axis on Figure \ref{fig:corb} and
increases in the same direction as the orbit's true anomaly.
To make the convention for this angle clear, in Figure \ref{fig:pan_sil} 
we show our impact or accretion angle $\theta_a$
with a labelled silhouette (based on Figure 5 by
\citealt{thomas20}) of Pan's equatorial ridge as seen from the north. 
W longitude in degrees is labelled in Figure \ref{fig:pan_sil} and on the right axis in the top
panel of Figure \ref{fig:ang}.
With convention of zero longitude on the moon's sub-Saturn side, an impact angle of 0 corresponds
to a longitude of $180^\circ$ W.  
Western longitude on the moon increases in the clockwise and opposite direction as our impact angle. 
Figure \ref{fig:ang} only shows particles originating exterior to the moon's orbit.
Particles originating from inside the moon's orbit exhibit similar phenomena, except
the impact angle should be shifted by $180^\circ$.

The tangential velocity component in the bottom panel of Figure \ref{fig:ang} 
is shown in units of Hill velocity, $v_H \equiv R_H n$
where $n$ is the mean motion of the moon.
Hill velocities in m/s are listed for the three moons in Table \ref{tab:computed}.
The tangential velocity component is computed from the velocity difference between test
particle and moon center and uses the moon center as origin to specify radial
and tangential directions.
The tangential velocity component times the radius gives the spin angular momentum per
unit mass of accreting material that impacts the moon.  
Figure \ref{fig:ang}'s bottom panel shows that the angular momentum per unit mass of accreting material
is of order the Hill velocity times the Hill radius, $v_H R_H$. 
This size-scale is consistent with the estimate for the torque due to accretion we made 
in section \ref{sec:torque_acc} (Equation \ref{eqn:Tacc}).  
The specific value of spin angular momentum  depends on the initial test particle 
  orbit semi-major axis
and at a lesser sensitivity, the location of the moon surface.
In tidal lock, the spin angular rotation rate of the moon on our plots is counter-clockwise.
Trajectories at larger initial semi-major axis offset $|\Delta a| $ 
have negative tangential velocity component $v_\theta$ and 
in the absence of tidal alignment, 
the accretion stream would slow the moon's rotation.  
The opposite is true for the test particles 
with smaller values of $|\Delta a|$ and these would
increase the moon's spin rate.

In summary, integration of initially circular orbits on first close approach shows that only particles 
with orbital semi-major axes satisfying $1.7 \lesssim |\Delta a|/R_H \lesssim 2.5$ 
are likely to impact a moon.   We confirm the range found previously through
numerical integration  \citep{karjalainen07} (see their Figure 1). 
Particles can impact the moon's surface at almost
any longitude on the moon's equator.  The impact longitude and angular momentum per unit mass
of accreting material depend on particle orbit semi-major axis and at a lower sensitivity
on the radius of the moon's surface within the Hill sphere.
 
\begin{figure*}   
\ifpreprint
\centering\includegraphics[width=6.0in, trim={14mm 0mm 0mm 0mm},clip]{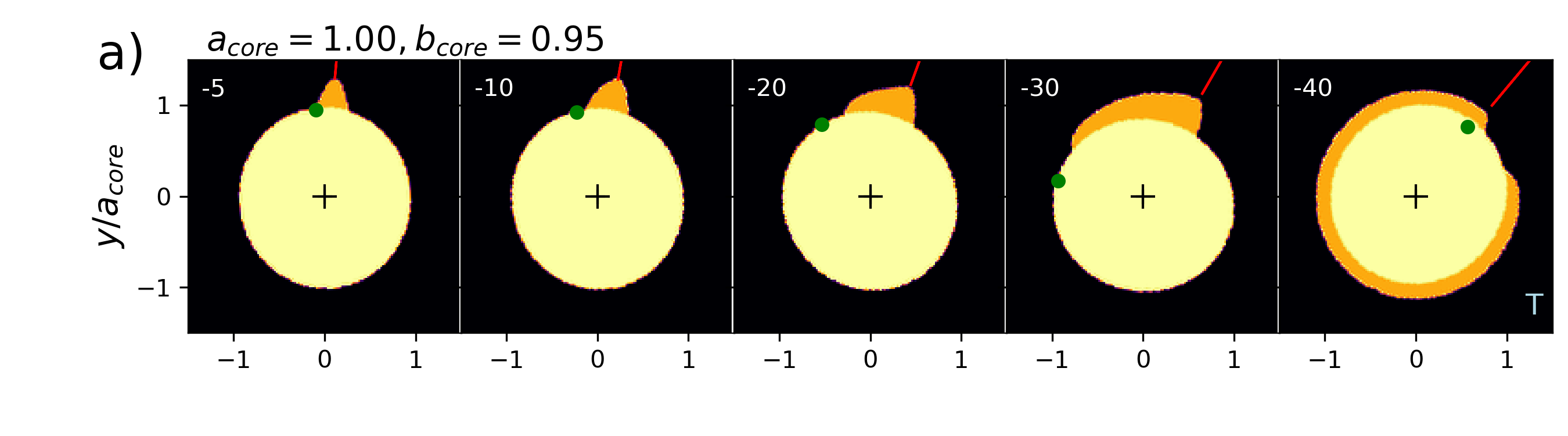} 
\centering\includegraphics[width=6.0in, trim={14mm 0mm 0mm 0mm},clip]{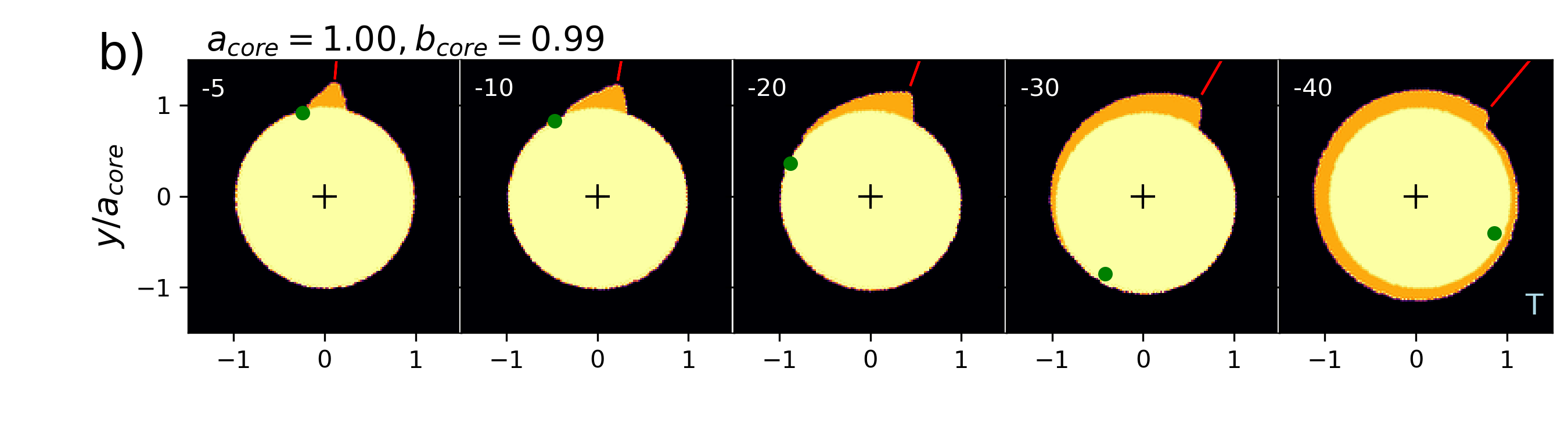} 
\centering\includegraphics[width=6.0in, trim={14mm 0mm 0mm 0mm},clip]{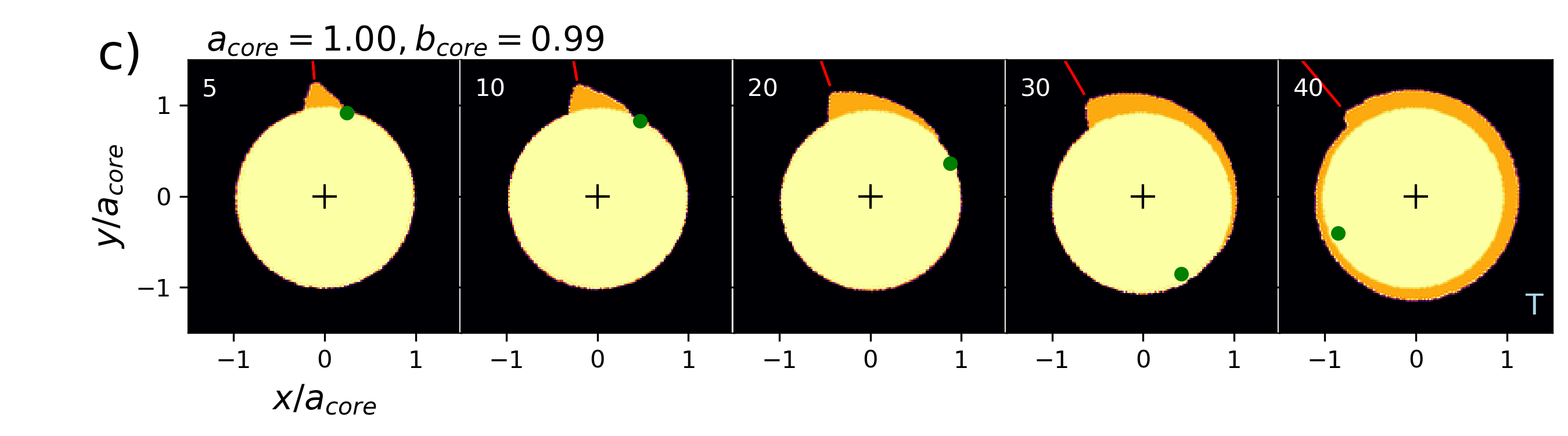} 
\else
\centering
\includegraphics[width=5.0in, trim={14mm 0mm 0mm 0mm},clip]{seriesthm5.png} 
\includegraphics[width=5.0in, trim={14mm 0mm 0mm 0mm},clip]{seriesthm1.png} 
\includegraphics[width=5.0in, trim={14mm 0mm 0mm 0mm},clip]{seriesth1.png} 
\fi 
\caption{The distribution of accreted mass for a tidally locked moon.
Each panel shows accretion at a different accretion angle $\theta_a$ with this angle in degrees 
labelled on the top left corner of each panel.  The accretion stream is shown with a red line segment 
pointing to  the impact point.
A `T' on the lower right of each row,
signifies that 
the body remains tidally locked during
accretion. However variations in body shape cause the moon to tilt slowly. 
The initial top side of the moon is labelled with a green dot. 
The tidal field is oriented so the moon's largest principal body axis 
remains aligned along the vertical direction.
The moon's core is shown in yellow.  The accreted material is shown in 
orange.  The center of mass is shown with a plus sign.  The axes are in units of 
the initial body semi-major axis 
which is initially oriented upward.    These accretion simulations were done in 2 dimensions.
a) The core's initial semi-axes are $a_{\rm core}=1.00$  and $b_{\rm core} = 0.95$ and the accretion angles
are negative.  The  slopes (derivative of radius w.r.t. longitude) 
of the lobes on Pan's equatorial ridge appear consistent with this sign.
b) The core's initial semi-major and minor axes  are 1.00  and 0.99 and the accretion angles
are negative.  
The angular shift in body orientation is higher than in a) due to the core's rounder shape.
c) The core's initial semi-axes  are 1.00 and 0.99 and the accretion angles
are positive. 
  \label{fig:seriesth}}
\end{figure*}

\subsection{Accretion while tidally locked and in a circular orbit}
\label{sec:acc_lock}

We consider a moon core, in a circular orbit, that is accreting at a rate $\dot M < \dot M_{cr}$
so the body remains tidally locked.   We assume that the moon accretes via
a narrow accretion stream that hits the moon 
 at a particular impact angle, $\theta_a$, on the moon surface.  If this angle differs
 from the principal body axes, then as mass piles up on the moon surface,
the orientation of the principal body axes, with respect to the core, will vary. 
  The moon would  tidally realign
so that its long principal body axis remains on the moon/planet line. 

Our accretion or impact angle  is zero, $\theta_a = 0 $,  when accretion flows directly through the L2 Lagrange point onto the moon and downward
 on Figure \ref{fig:corb}.
This accretion angle increases in the counter-clockwise direction, since
it is essentially the same thing as  the impact angle
discussed above (see Figure \ref{fig:pan_sil}).
  
To  explore numerically how accretion causes shape and body alignment changes, we mark a two-dimensional density array with ones and zeros depending upon whether mass is present in the pixel or not.  
The array pixels record density at locations in a Cartesian coordinate system for a two-dimensional body
in the equatorial plane.
The ones correspond to a uniform density material and zeros are empty space outside both core and accreted material.

The moon core is initialized by filling an oval region in the center of the density array with ones.  The orientation of the body (and our density array) in the frame rotating with the moon's orbit is given with an orientation angle $\phi$.   
Mass is consecutively added to the body surface along the impact angle $\theta_a$ to individual pixels in the array on the body surface.   Mass is accreted by choosing a pixel that contains a zero value
and changing it to one. 
We create a cost function for each pixel that increases as a function
of angle from $\theta_a$ and with radius.  The unfilled empty pixel with the minimum cost
function is chosen during each mass accretion event.   The cost function we used is the product
of an exponential function of radius and an exponential function of distance from the line 
at angle $\theta_a$ that begins at the origin.  

After a single pixel is filled, we recompute the center of mass position and the orientation of the body principal axes.
Our procedure for computing the orientation of the long principal body axis is described in \ref{ap:mom}.
We rotate the body about its center of mass by updating $\phi$ so that the body's longest principal axis remains aligned vertically.  This rotation maintains tidal alignment.  
We recompute the center of mass and $\phi$ for each
parcel of mass that is added to the body.  
Figures or rows of figures where tidal alignment is maintained
during accretion are marked a blue 'T' in one of the corners of the figure.

We illustrate body shapes and orientations arising from accretion along 5 different accretion 
angles in Figure \ref{fig:seriesth}. 
The axes are shown in units of the moon's initial core semi-major axis. 
The initial core is shown in yellow and the accreted material
is shown in orange.  A green dot shows a position on the core that was initially oriented 
on the anti-planet side (upward as in our previous figures). 
Each panel in Figure \ref{fig:seriesth} shows accretion at a different angle $\theta_a$ that is  
labelled in degrees on the top left side of the panel.  The angle of accretion is 
marked with a short red segment outside the body.

For small values of impact angle $\theta_a$, accreted mass continues to pile up near the impact point.
The accreted mass increases the longer principal body axis. 
The more mass that is added, the less the orientation angle of the core varies. 
This trend is illustrated with a simple model in \ref{ap:mom}  (see equation \ref{eqn:deltaphi}).  
This gives a triangular wedge shape to the accreted material.  
The wedge shape resembles the shape of a single lobe of
Pan's equatorial ridge.
If $|\theta_a| \gtrsim \pi/4$ then the accreted mass tends to increase the mass distribution along the smaller principal body axis of the body (see equation \ref{eqn:deltaIxxyy}).  The body becomes rounder 
and the angular shift caused
by accretion is increased (equation \ref{eqn:deltaphi}). 
The body continues to rotate, accumulating mass evenly around 
its equatorial ridge.    

The lobes on Pan's equatorial ridge seem to increase in thickness and radial extent as the western longitude (on the equator) increases (see Figure \ref{fig:moons}), though this trend is not evident in the thinnest
and most extended part of the ridge and 
from the silhouette shown in Figure \ref{fig:pan_sil}.  
We can characterize Pan's lobes with a slope that is the derivative of their radial extent with respect
to longitude.
The wedges in our accretion model with accretion angle 
$\theta_a < 0$, as shown in Figure \ref{fig:seriesth}a, b. 
have the same slope as the lobes of Pan.
The wedges have the opposite slope
if the accretion angle $\theta_a >0$, as shown in Figure \ref{fig:seriesth}c.

The rate the body core orientation angle changes while accreting 
depends on both core shape and on the accretion angle.   If the core is elongated
then more mass must be accreted to cause the same degree of rotation during accretion.
This is illustrated with a comparison between  Figure \ref{fig:seriesth}a,  and b.
The initial cores in Figures \ref{fig:seriesth}b,c  have semi-axis lengths 1.00  and 0.99
whereas those in  \ref{fig:seriesth}a are initially more elongated with semi-axis lengths 1.00  and 0.95.

The simple accretion model explored here gives two types of morphology for
material accreted on an equatorial ridge, single wedges or uniform accretion covering 
the entire equator. If accreting material came from both inside and outside the moon's
orbit, the accreted material would be symmetrically distributed, with material
coming from the planet side shifted by $180^\circ$ from that accreted from the
anti-planet side.  The model does not give polygon-shaped equatorial ridges, like Pan, or 
ridges with a depression on one side of the moon, like Atlas.  Because accretion has
been restricted to the plane we also do not see ridges at different latitudes like those on Daphnis.

In summary,  a tidal accretion model that takes into account body shape changes
while maintaining tidal alignment and at a single accretion angle can 
give wedge-shaped or sloped lobes along an equatorial ridge if the accretion angle is low
with respect to the moon/planet line.
However, the model only illustrates single wedges or uniform accretion covering 
the entire equator.   More complex models are required to match the morphology
of Pan, Atlas or Daphnis.

\begin{figure} 
\centering\includegraphics[width=3.2in, trim={0mm 0mm 0mm 0mm},clip]{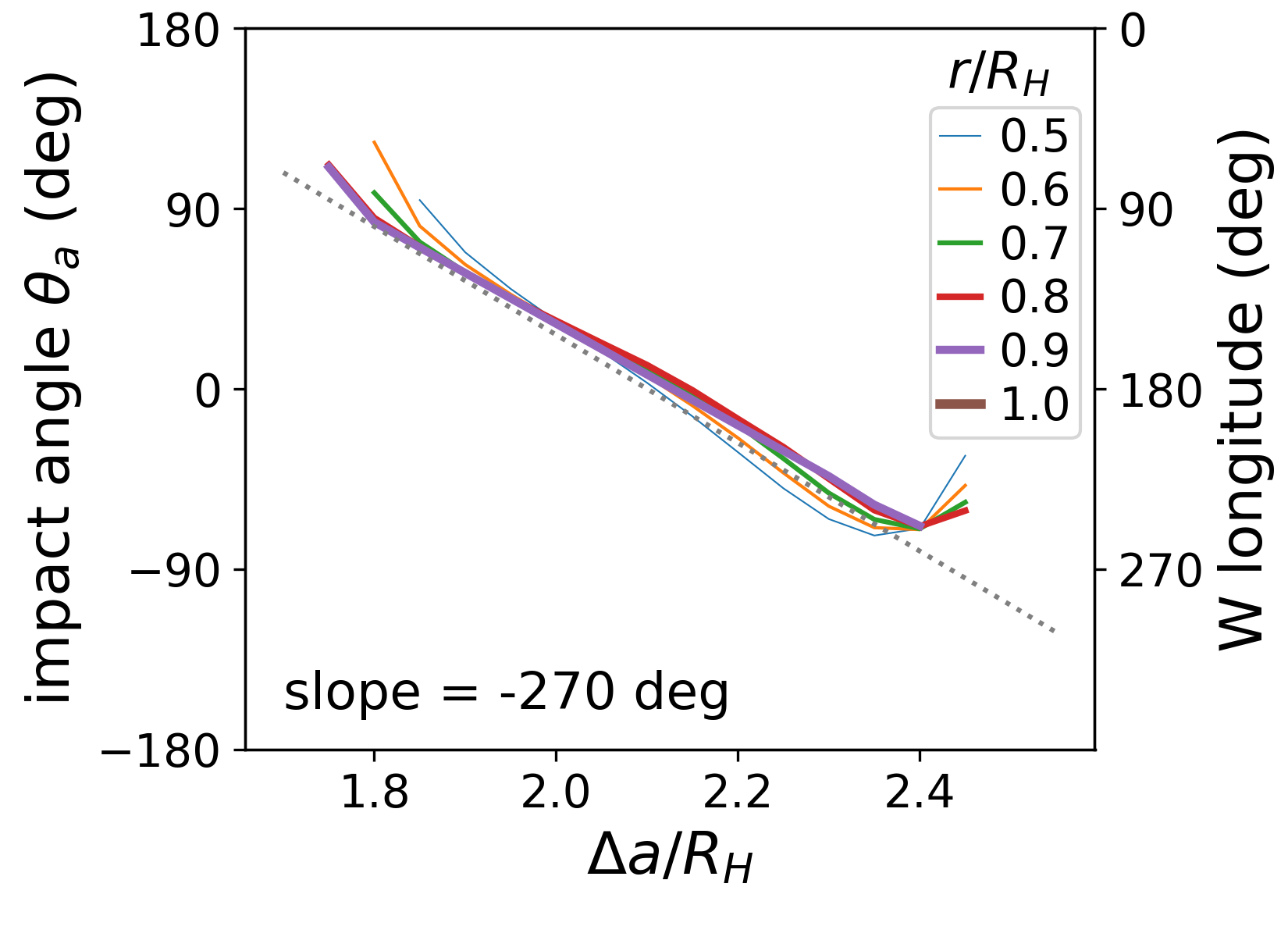} 
\caption{Impact or accretion angle $\theta_a$  in degrees as a function 
of initial test particle orbital semi-major axis distance from that of the moon $\Delta a/R_H$.  These are
computed from the same initially circular orbits external to a moon in 
a circular orbit that are shown in Figure \ref{fig:corb}.
We have plotted the impact angle at five different surface radii from the moon center. 
The different lines represent the different surface radii and they are given in the key in units of the Hill radius $R_H$.
The impact angle is only weakly dependent on the radius of the surface but is strongly dependent 
on the test particle's initial orbital semi-major axis.
The grey dotted line is given in equation \ref{eqn:slope}. 
\label{fig:a_theta}
}
\end{figure}

\begin{figure*} 
\ifpreprint
\centering\includegraphics[width=7.0in, trim={10mm 0mm 0mm 10mm},clip]{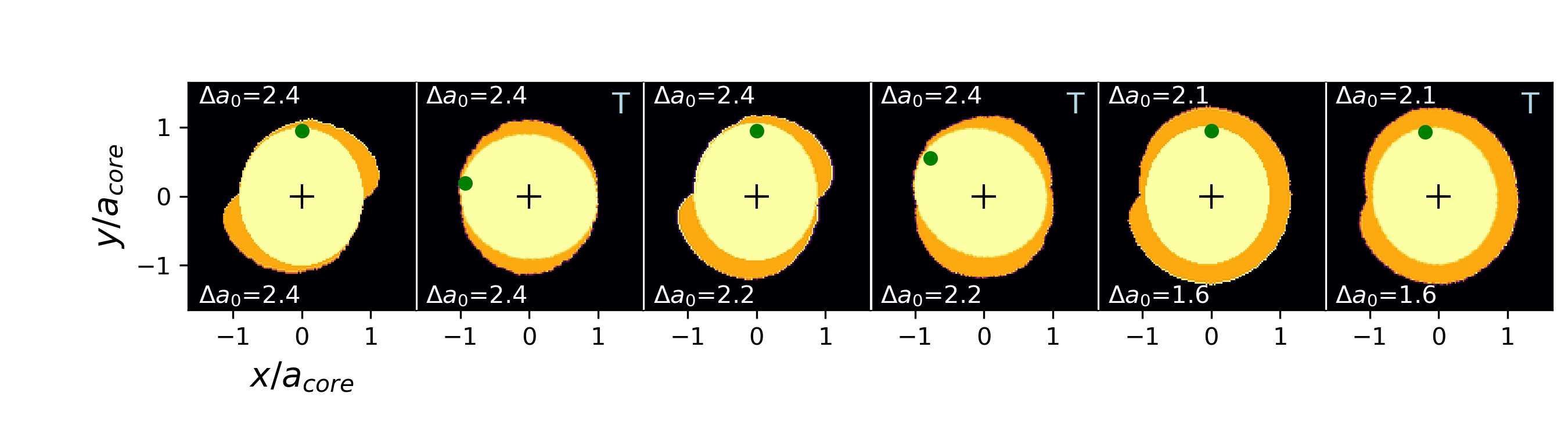}
\else
\centering\includegraphics[width=5.6in, trim={10mm 0mm 0mm 10mm},clip]{sigprof.png}
\fi
\caption{
The distribution of accreted mass on a moon's equatorial ridge where 
accreted material originates from the outer and inner gap edges with
mass surface density profiles that are described with equation \ref{eqn:profile}.  
This figure is similar to Figure \ref{fig:seriesth} except accreted material 
originates from    a range of orbital semi-major axes and impact angles.  
The outer
gap edge has edge location $\Delta a_0$ in units of $R_H$ labelled on the top of each panel
and the inner edge has  $\Delta a_0$ in units of $R_H$  labelled on the bottom of each panel. 
Only the panels labelled 'T' on the top right remain tidally locked during accretion.
The others do not allow the body to rotate. 
The accreted material is shown in 
orange and the core is shown in yellow.  
The center of mass is shown with a plus sign.  The plot axes are in units of 
the initial body semi-major axis 
of the moon which is initially oriented vertically.   
The core's initial body semi-axes  are $a_{\rm  core} = 1.00$  and $b_{\rm  core} = 0.90$.  
 \label{fig:sigprof}}
\end{figure*}

\subsection{Accretion from a gap edge}
\label{sec:c_edge}

Up to this point we have considered accretion from material from a narrow ring of disk 
material that has a single orbital semi-major axis and giving an accretion stream
that is confined to a single impact angle. 
Gravitational encounters with the moon usually push ring material away from
the moon \citep{goldreich82,weiss09}.  
The gap widens until torques driven by spiral density waves 
balance those associated with viscous spreading 
\citep{borderies82,borderies89,lissauer81,goldreich82,porco05,tajeddine17a}.
Right now the Keeler gap hosting Daphnis and the Encke gap hosting Pan are sufficiently
wide that there is little or no accretion onto these moons.  However, the gaps might
have been narrower at some time in the past, allowing accretion.
Models taking into account spiral density wave driven torque and diffusion in the 
gap edge have predicted gap edge surface density profiles  \citep{gratz19b}. 
The gap edge could display small scale structures, such as 
the few hundred meter length wispy features near the Keeler gap edge \citep{porco05,tajeddine17a}.  
In both settings, the surface density profile (averaged over ring longitude if wisps are present) drops
significantly over a distance $d_{\rm edge}$ that is of order a Hill radius.

If the region allowing accretion overlaps the region where the surface density profile drops,  
then we expect more mass is accreted from impact angles arising
from the denser and more distant regions of the disk.
To illustrate the dependence of impact angle on initial semi-major axis
we plot in Figure \ref{fig:a_theta} the impact angle at 5 different radii from
the body center as a function of initial orbital semi-major axis offset $\Delta a$ from
that of the moon. We have only plotted initially external orbits, with $\Delta a>0$, using 
the same orbits as shown in Figure \ref{fig:corb}.
The impact
angle is weakly dependent on the radius of the surface but is strongly dependent 
on the test particle's initial orbital semi-major axis.
The impact angle is approximately linearly dependent on initial test particle orbital 
semi-major axis offset $\Delta a$ and is reasonably well described by 
\begin{equation}
\theta_a(\Delta a) \approx - 270^\circ \left( \frac{\Delta a}{R_H} - 2.1 \right). \label{eqn:slope}
\end{equation}
We show this relation as a dotted grey line on Figure \ref{fig:a_theta}.

What slope in a gap edge would give lobes similar to those present
on Pan's equatorial ridge?    If the lobes vary by a factor of 2 in their linear
mass density (mass per unit distance along the equator) across $90^\circ$ in 
equatorial longitude, then the slope of equation \ref{eqn:slope}
gives a distance  $d_{\rm edge} \sim 0.3 R_H$  over which the surface density
drops.  This distance seems reasonable 
compared to analytical models of gap density profiles \citep{gratz19b}.

Figure \ref{fig:sigprof} illustrates some possible equatorial ridge morphologies
that take into account the mass surface density profile in both inner and outer edges.
We start with a surface density profile $\Sigma(\Delta a)$ describing  edges with a Gaussian form
\begin{equation}
\Sigma(\Delta a) = \begin{cases} 
\exp\left( - \frac{(|\Delta a| -\Delta a_0)^2}{2 R_H^2 \sigma_h^2} \right)
    & \mbox{if } |\Delta a| < \Delta a_0 \\
1   & \mbox{otherwise}
\end{cases} \label{eqn:profile}
 \end{equation}
where $\Delta a_0$ describes the edge location
and $\sigma_h$ is a dispersion that determines how fast the profile drops near $\Delta a_0$.
For equally separated
values of $\Delta a$, we add mass particles onto our density array
(as described in section \ref{sec:acc_lock}) at impact angle 
that is computed using Equation \ref{eqn:slope} and is sensitive to $\Delta a$.
Mass particles are added onto the body's 
surface with probability set by $\Sigma (\Delta a)$. 

Equation \ref{eqn:slope} is used to relate $\Delta a$ to the impact
angle and determines where mass is added to the modeled equatorial ridge.   
Only mass originating between $1.7 < |\Delta a/R_H| < 2.4$ is allowed
to accrete onto the body, as impacts don't occur outside this range.  
We  simultaneously accrete from both outer and inner
edges of a gap but we need not have the same
edge location or gap profile dispersions in both edges.  
As described in section \ref{sec:acc_lock}, the body can be continually
reoriented while it is accreting so as to remain as if it were tidally locked throughout 
accretion.

Figure \ref{fig:sigprof} shows 6 different equatorial ridge morphologies 
computed with gap profile dispersion $\sigma_h = 0.2$ and for different values of edge locations $\Delta a_0$
for inner and outer edges.  The panels are in pairs, with tidally locked accretion
models to the right of a similar model with body rotation held fixed during accretion.
The specific values for $\Delta a_0$ for the outer gap edge are written on the top
of each panel and on the bottom of each panel for the inner gap edge.
A gap profile gives 
 higher mass density at larger $|\Delta a|$.  Along with Figure \ref{fig:a_theta}, this 
implies that the linear mass density of an accreted lobe would 
 increase with increasing western longitude (corresponding to more negative impact 
 angle $\theta_a$).   This trend is evident in Figure \ref{fig:sigprof} for most of
 the modeled accreted lobes and 
 the mass per unit longitude of Pan's wedge-shaped lobes seem to increase in this direction.
 
If the gap profile remains constant  during the accretion era, then 
accretion from the edge of a gap while in a circular orbit would allow growth
of a single wedge-shaped lobe.    If the moon simultaneously accretes from
both outer and inner gap edges then two wedge-shaped lobes would accrete
onto an equatorial ridge.  If the outer gap edge is further away from the moon
than the inner gap edge (the two rightmost panels in Figure \ref{fig:sigprof})
then less mass is accreted on the moon's leading side.  An alternative
explanation for the divot on the leading side of Atlas's ridge 
might be a difference in inner and outer gap edge surface density profiles.
While Figure \ref{fig:sigprof} illustrates a variety of equatorial ridge shapes,
we don't see an equatorial ridge with multiple peaks or lobes like Pan's.

In summary, when accreting material is originally in a circular orbit and impacts
a moon in a circular orbit, there is a nearly linear relation between the
initial semi-major axis offset $\Delta a$ of the accreting material from the moon, 
and the impact angle on the moon's equator.  
A moon accreting from an outer gap edge would accrete more
mass at larger western longitude on its equator.  The slope of an accreted wedge would
be related to the slope of the surface density profile in the gap edge.
 If the outer gap edge is further away from the moon
than the inner gap edge then less mass is accreted on the moon's leading side,
 giving a possible alternative explanation for structure on Atlas's equatorial ridge.
However,  accretion onto a moon in a circular orbit  
from inner and outer gap edges does not seem to give an equatorial ridge
with multiple peaks or lobes like Pan's.

\begin{figure} 
\centering\includegraphics[width=3.3in, trim={0mm 0mm 10mm 0mm},clip]{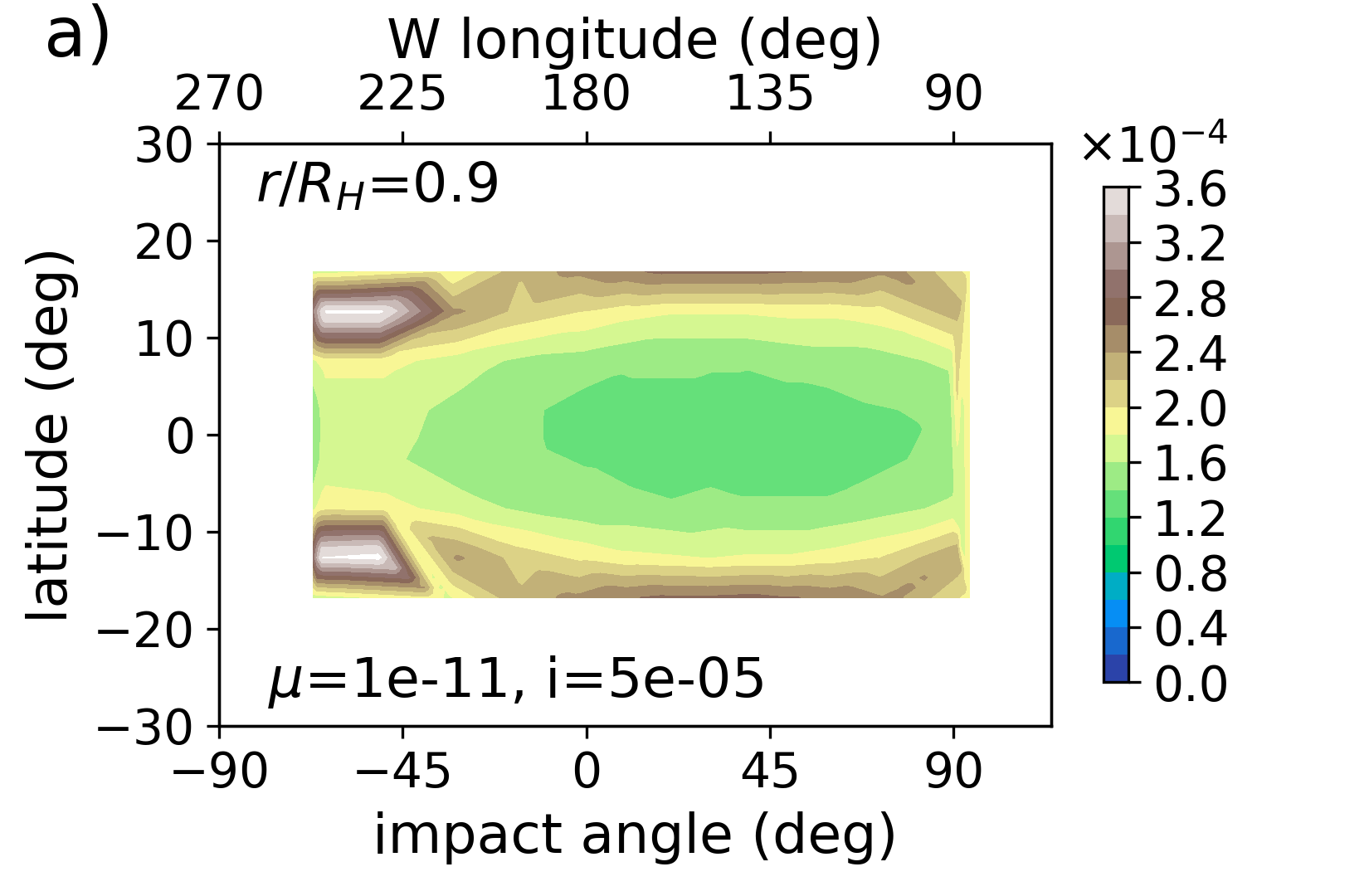}
\centering\includegraphics[width=3.3in, trim={0mm 0mm 10mm 0mm},clip]{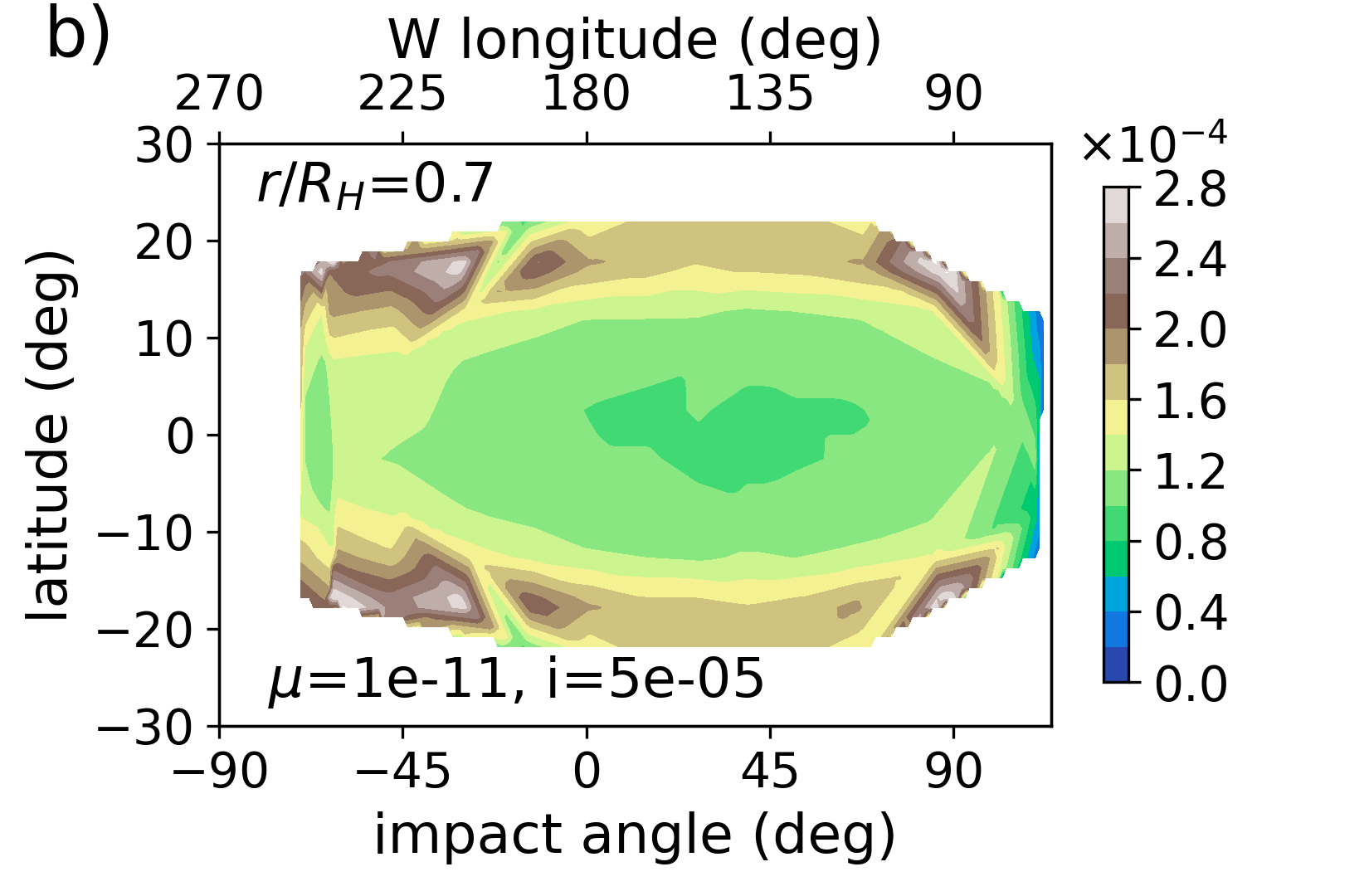} 
\caption{
Distribution of impact latitude and longitude on a moon by test particles
initially in zero inclination and circular orbits, external to a moon's orbit,
 that impact a moon on an inclined orbit.
The bottom axes shows our impact angle $\theta_a$  in degrees and the top axes
shows W longitude.  The vertical axis is latitude in degrees.  
The moon has mass ratio $\mu=10^{-11}$ and orbital inclination $i=5 \times 10^{-5} \approx 0.3 i_H$.
The colour map shows the fraction of test particles
per square degree that impacted at a particular location. 
a) Impacts are recorded  at $r = 0.9 R_H$ from the moon center of mass.
b) Impacts are recorded  at $r = 0.7 R_H$ from the moon center of mass.
 \label{fig:lat}}
\end{figure}

\subsection{Accretion onto a  moon on an inclined orbit}
\label{sec:inc}

\citet{charnoz07} suggested that the thickness of the equatorial ridge of
Pan and Atlas might be related to their orbital inclinations. 
We examine impact locations for test particles initially in circular orbits
that impact a moon that is on an inclined orbit with respect to the plane containing 
the test particles.
During the encounter, 
test particle trajectories are sensitive to the height of the moon, in its orbit, above or below
the plane containing the test particles (the ring plane).  
The Hill inclination  $i_H \equiv \frac{R_H}{a}$
is the inclination  required for the moon to travel a Hill radius $R_H$ above
the ring plane or equivalently Saturn's equatorial plane. 
At large inclination
($i>i_H$), impact with the moon is less likely because the distance between
particle and moon orbits can be larger than if both were in the same plane.
Particles that  impact the moon can impact over a wide range of possible latitudes on the moon.
Because the equatorial ridges of Pan, Daphnis and Atlas are thin, their  orbital
inclinations probably remained low during accretion of their equatorial ridges \citep{charnoz07}, 
with  $i<i_H$.
To explore this regime, we carried out test particle integrations for a moon with mass ratio 
$\mu=10^{-11}$ and 
with orbital inclination $i=5 \times 10^{-5}$ which is $\approx 0.3 i_H$.
For comparison, the inclination of Pan is probably less than its Hill inclination  (within errors),
whereas Daphnis' current inclination is 100 times its Hill inclination
and Atlas' inclination is 20 times its Hill inclination.
Due to its orbital inclination, 
Daphnis excites vertical waves on the edges of the Keeler gap \citep{weiss09}. 

Test particle trajectories near the moon depend on their initial orbital true longitude
because upon close approach, the moon can be at different heights above
or below the ring plane.  
We integrated $N_{\rm test} = 400$ test particle orbits for each  initial test particle 
semi-major axis, $\Delta a$, but each particle in the set is initially at a different orbital longitude.
The separation in orbital longitude between consecutive initial test particles  
is $\frac{2 \pi \Delta n}{N_{\rm test}}$ where 
$\Delta n = n' - n  \approx n \frac{3}{2} \frac{\Delta a}{a}$ is the difference between test particle
and moon mean motions.  This separation ensures that we sample the entire
range of possible close encounter behaviors.  

Figure \ref{fig:lat} shows the distribution of impact locations, computed 
from orbits with initial semi-major axes $\Delta a/ R_H \in 1.75 $ to 2.5 with a spacing of 0.05. 
The impact angle or W longitude is set by the initial orbit semi-major axis $\Delta a$.
At a particular $\Delta a$, 
since the inclination is low, the test particle impact angle is within a few degrees of
that computed when the moon orbital inclination is zero 
(see Figure \ref{fig:a_theta}). 
Figure \ref{fig:lat}a  shows impact surface density maps for orbits that impact a spherical surface 
 at a radius $r = 0.9 R_H$ from the moon center.  
Figure \ref{fig:lat}b is similar except at $r = 0.7 R_H$.  The latitude and longitude
distributions are wider at the smaller surface radius due to the gravitational acceleration of the moon itself.
This is consistent with the higher impact velocities seen in the lower panel of Figure \ref{fig:ang}.

Figure \ref{fig:lat} shows that fewer impacts hit  
 near the equator than at latitudes of $10$--$20^\circ$.  
The higher latitude impacts are test particles 
 deflected vertically when the moon was cresting above or below the ring plane. 
 We confirm  
 the peaked impact latitude distribution shown in the simulations by \citet{charnoz07}.

Pan's and Atlas' equatorial ridges are not U-shaped.  They don't  have a divot on 
their equator and bulges above and below the equator, as would be suggested by
Figure \ref{fig:lat} and Figure 2A by \citet{charnoz07}.   
We have not explored scattering or rolling on the surface after impact or 
inter-particle collisions during accretion.
Latitude distribution would also 
be sensitive to orbital eccentricity.
Models that include additional 
 processes are probably needed to explain the thickness distribution of Pan's 
equatorial ridge.  However, some of the fine features present on these ridges
are likely to be related to the orbital inclination of the moon during
accretion and the inclination of the particles that were accreted.

In summary, we have examined the latitude distribution of impacts for accretion of a moon
in a circular but inclined orbit by ring material that is initially in a circular orbit.
As was true for coplanar moon and particle circular orbits, the longitude of impact is dependent on the initial test particle semi-major axis.
At low inclination, $ i \lesssim i_H$, test particle impact angle is similar to the impact angle when the moon and test particle orbits are coplanar.
We integrated impacting test particle orbits for a moon with mass ratio
$\mu = 10^{-11}$ and orbital inclination
$i = 5 \times 10^{-5} \approx 0.3 i_H$ and found that the latitude distribution of impacts is U-shaped, 
confirming the simulations by \citet{charnoz07}, with more particles impacting at latitudes of 10--$20^\circ$ than on the equator.  This implies that orbital eccentricity or additional processes, such as scattering on the surface, collisions in the accretion stream, or mass redistribution are required to account for ridge thickness and latitude distribution.  

\begin{figure}  
\ifpreprint
\centering\includegraphics[width=3.0in, trim={0mm 0mm 0mm 0mm},clip]{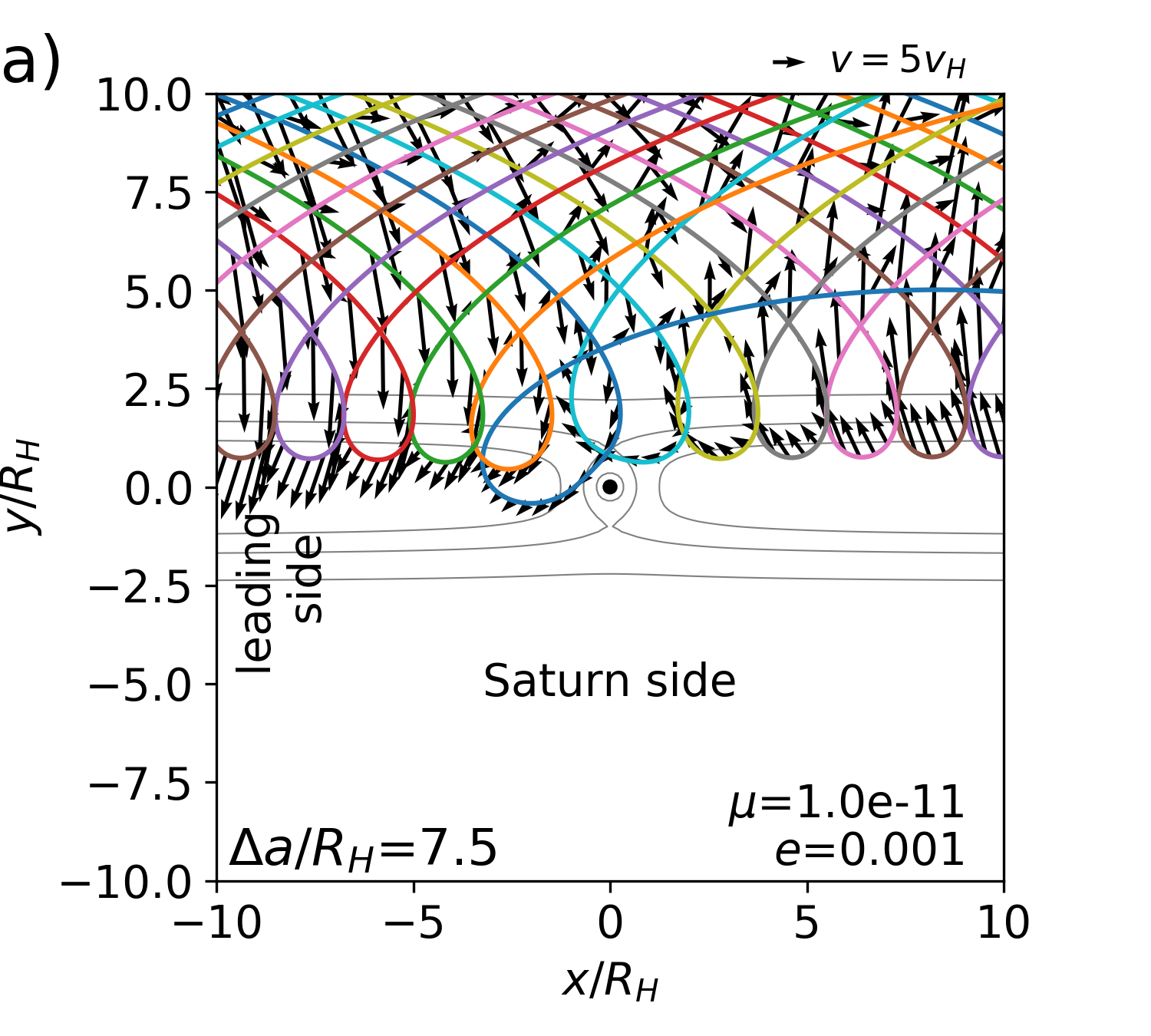}
\centering\includegraphics[width=3.0in, trim={0mm 0mm 0mm 0mm},clip]{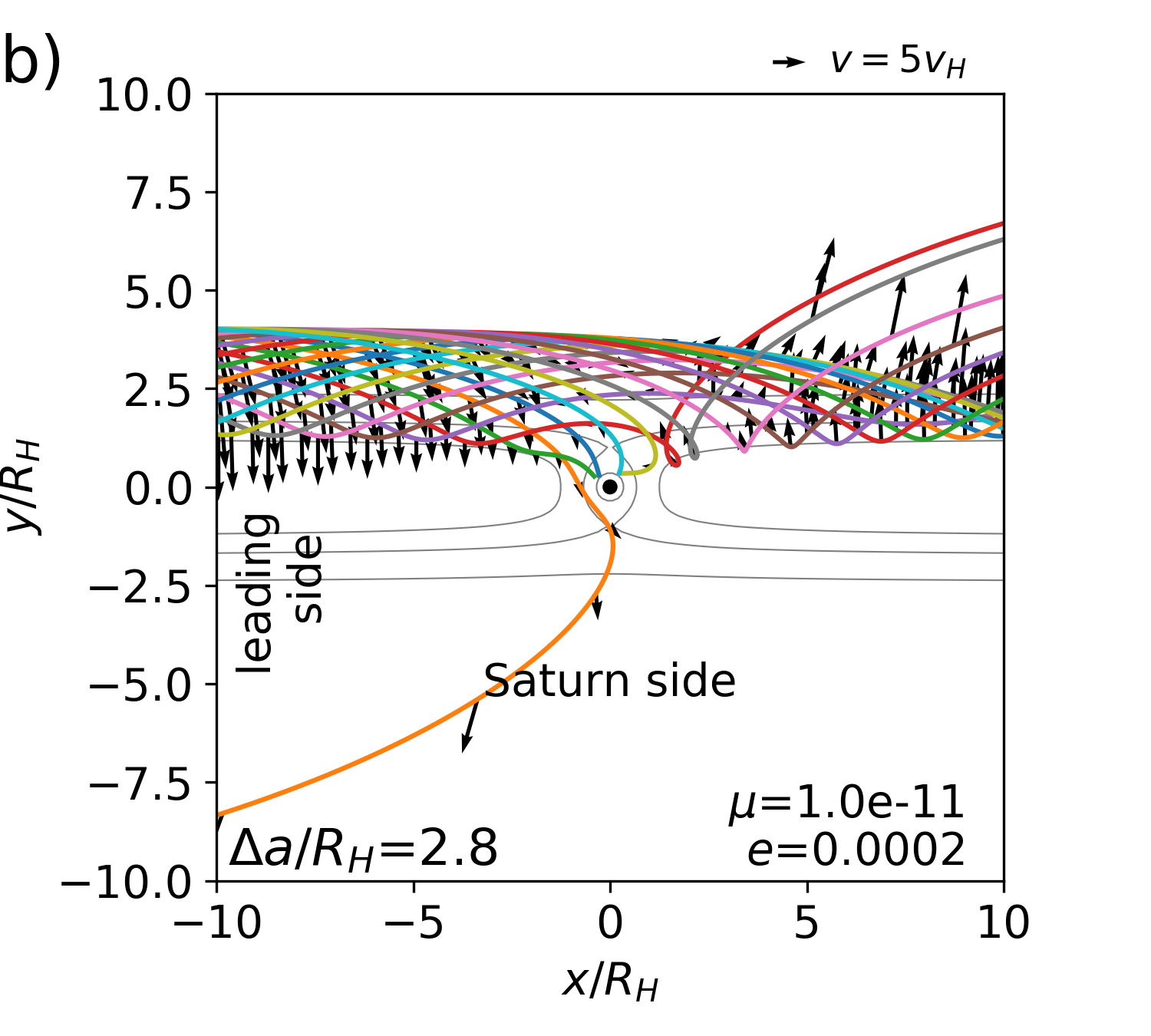} 
\else
\begin{tabular}{cc}
\centering\includegraphics[width=3.0in, trim={0mm 0mm 0mm 0mm},clip]{a75_large.png} &
\centering\includegraphics[width=3.0in, trim={0mm 0mm 0mm 0mm},clip]{b28_large.png} 
\end{tabular}
\fi
\caption{
Test particle orbits near a moon in an eccentric orbit.
These figures are similar to Figure \ref{fig:corb} except the moon has an orbital eccentricity.
The mass ratio  $\mu = 10^{-11}$ is similar to that of Pan.
The planet facing side of the moon is on the bottom and the leading  side of the moon 
is on the left.
Each orbit has the same initial orbital semi-major axis but begins at a different orbital true
longitude. 
The orbits are shown in a frame that puts the moon center at  the origin
and the planet on the negative $y$ axis.
Black velocity vectors show the difference between test particle and moon velocity.
The scale of these vectors is shown with an arrow on the upper right that
has length 5 $v_H$ where $v_H$ is the Hill velocity
$v_H = nR_H$.
The test particles are initially on circular
orbits  outside the orbit of the moon with initial semi-major axis
$\Delta a$ shown on the bottom left of the plot.  When the moon
is at apocenter, the  test particles are at the bottom of a loop. 
Because the angular rotation rate of the moon slows at apocenter, the test particles
can approach the moon from the right, rather than the left. 
a) The moon's eccentricity is $e=0.001 \approx 7 e_H$.
b) The moon's eccentricity is $e=0.0002 \approx 1.4 e_H$.
The size of the loops depends on the moon's eccentricity and initial test particle orbit semi-major axis
offset $\Delta a/R_H$.
  \label{fig:large}}
\end{figure}

\begin{figure*} 
\ifpreprint
\centering
\includegraphics[width=7.0in, trim={0mm 10mm 0mm 10mm},clip]{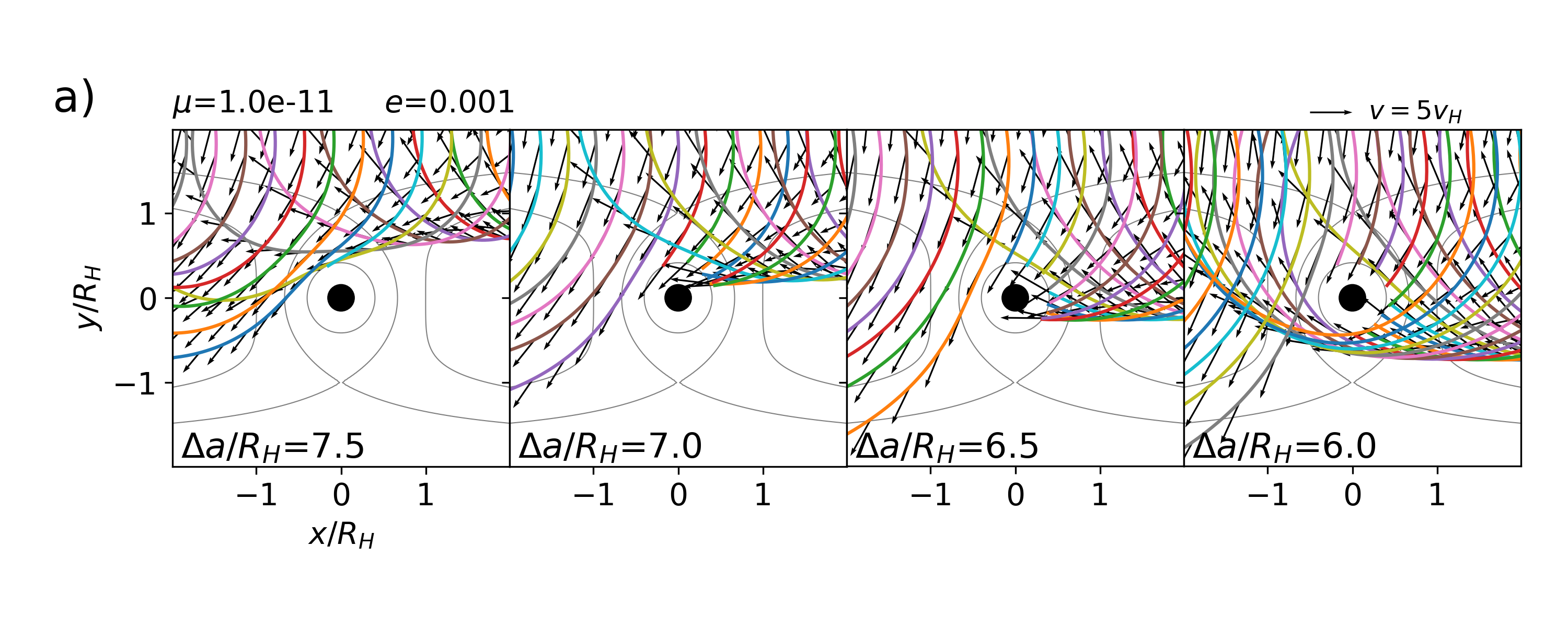} 
\includegraphics[width=7.0in, trim={0mm 10mm 0mm 10mm},clip]{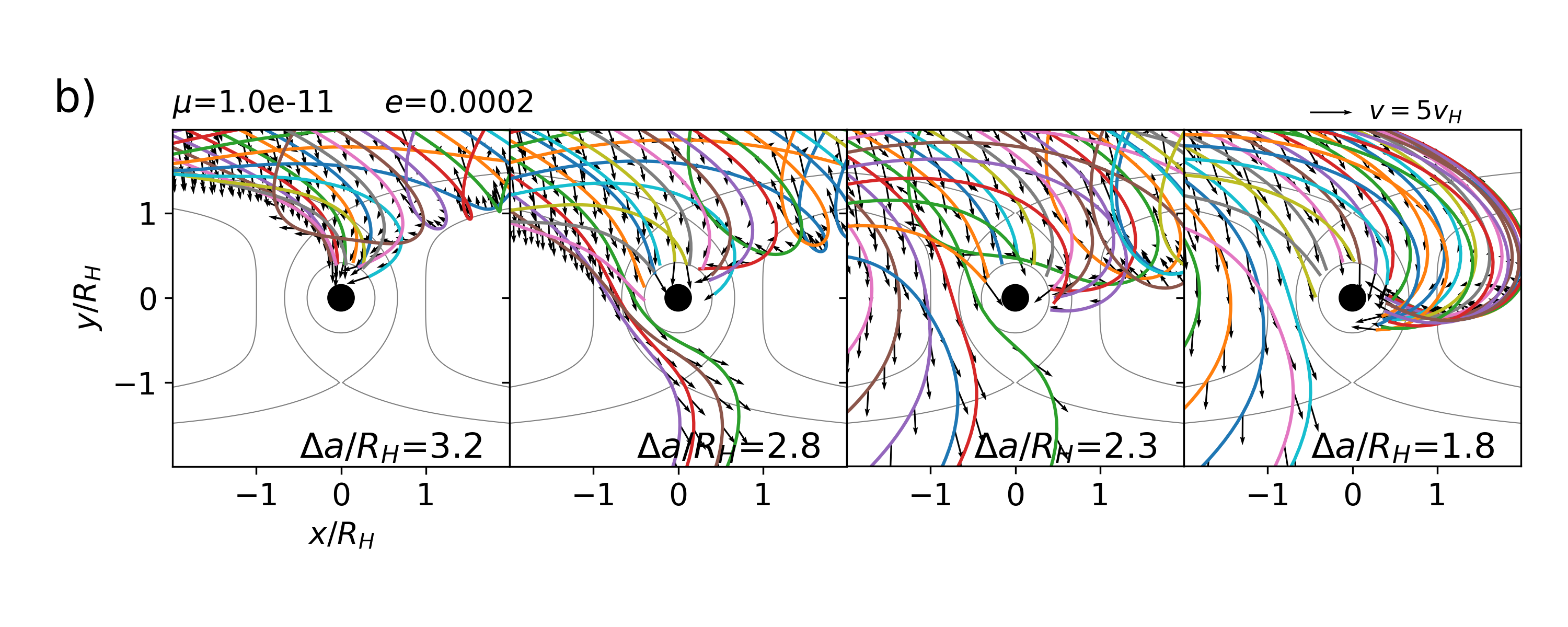} 
\else
\centering
\includegraphics[width=6.0in, trim={0mm 10mm 0mm 10mm},clip]{a_eorb.png} 
\includegraphics[width=6.0in, trim={0mm 10mm 0mm 10mm},clip]{b_eorb.png} 
\fi
\caption{
Test particle orbits near a moon of mass ratio $\mu=10^{-11}$ and  in an eccentric orbit.
Each panel is similar to Figure \ref{fig:large} except the axes span a smaller range.
Each panel shows a different initial semi-major axis for the test particle orbits, with
$\Delta a/R_H$ labelled on the lower left of each panel.
In each panel, the different test particle orbits are begun in circular orbits at different but
evenly spaced orbital true longitudes.
a) The moon has orbital eccentricity  $e=0.001\approx 7.1 e_H$. 
b) Similar to a) except $e=0.0002 \approx 1.4 e_H$ and with different values for the initial test orbit semi-major axis offsets $\Delta a/R_H$.
  \label{fig:eorb}}
\end{figure*}

\section{Test particles that impact a moon in an eccentric orbit}
\label{sec:eorb}

\citet{charnoz07} showed that the longitude distribution of accreting ring material onto a moon 
depends upon the moon's orbital eccentricity.  In this section, we look at impact angles for test particles,
initially on circular orbits, 
that approach a moon that is on an eccentric orbit.
For a moon on a circular orbit, the morphology of the approaching test
particle orbits in units of the Hill radius
is not strongly dependent on moon-to-planet mass ratio, $\mu$.  
However if the moon is on an eccentric 
orbit, then an additional spacial scale enters the problem, the radial distance travelled
between pericenter and apocenter.    If the radial distance travelled in the orbit is less than a Hill
radius, the impacting orbits should be similar to those impacting 
a moon in a circular orbit.  The Hill eccentricity 
\begin{equation}
e_H \equiv \frac{R_H}{a} \label{eqn:e_H}
\end{equation}  
is the orbital eccentricity required for the moon to travel $2R_H$ radially between orbit pericenter
and apocenter.  If the orbital eccentricity $e \ll e_H$, 
we expect the impacting orbits would be similar to those discussed in section \ref{sec:corb}.
We list the Hill eccentricities for Pan, Daphnis and Atlas in Table \ref{tab:computed}.
Currently Pan's orbital eccentricity is below its Hill eccentricity, whereas
Daphnis has an eccentricity similar to its Hill eccentricity and Atlas's eccentricity exceeds its
Hill eccentricity.

The distance from the edges of the Encke gap to Pan in units of Pan's orbital semi-major axis are $\Delta a/a =1.2 \times 10^{-3}$ and in units of Hill radius are $\Delta a/R_H = 8.45$.
The distance from the edges of the Keeler gap to Daphnis in units of Daphnis's semi-major axis
is $\Delta a/a= 1.5 \times 10^{-4}$ and in units of Hill radius $\Delta a/R_H = 4.3$.
The outer edge of the A-ring is at a semi-major axis of $a \sim  136,769 $ km 
approximately at the location of 
 the 7:6 inner Lindblad resonance with Janus \citep{porco84,spitale09}.
The distance of Atlas from this edge is currently $\Delta a/a = 6.5 \times 10^{-3}$ 
or $\Delta a/R_H = 43$.
As the distances to the ring edges in units of Hill radius are greater than 1, an eccentricity $e>e_H$ would be required
for the moons to accrete ring material from gap edges.  However, in the past,  
ring edges could have been at different distances from the moons and 
the moon eccentricities could have been higher. 
 

To characterize the regime where moon orbital eccentricity exceeds the Hill
eccentricity $e>e_H$,
 we examine impacting orbits for a moon
with mass ratio $\mu = 10^{-11}$ similar to that of Pan, and with an orbital eccentricity large enough
that Pan can graze  the edge of the Encke gap, $e = 0.001 \approx 7.1e_H$.    We also examine 
impacting orbits at a moon eccentricity of $e=0.0002$ which is $\approx 1.4 e_H$ in terms 
of Pan's Hill eccentricity
and about 14 times Pan's current eccentricity.  These integrations 
characterize the regime with eccentricity $e \sim e_H$.
With a moon on an eccentric orbit, 
test particles are initially in circular orbits to mimic originating from a dense collisional ring edge.
Test particle orbits for $e = 0.001$ and 0.0002 
are shown in Figure \ref{fig:large} and at a smaller scale in Figure \ref{fig:eorb}.
Orbit positions have been rotated and shifted so that the moon remains at the origin and the planet remains along the negative $y$-axis.

Test particle trajectories near the moon depend on their initial orbital true longitude
because upon close approach, the moon can be at different radii from the planet.
As in the case of the inclined moon (section \ref{sec:inc}), 
we integrated $N_{\rm test} = 400$ test particle orbits for each  initial test particle 
semi-major axis, where each particle in the set is initially at a different longitude.
The separation in orbital longitude  between consecutive initial test particles  
is $\frac{2 \pi \Delta n}{N_{\rm test}}$ where 
$\Delta n = n' - n  \approx n \frac{3}{2} \frac{\Delta a}{a}$ is the difference between test particle
and moon mean motions.  This separation ensures that we sample the entire
range of possible close encounter behaviors.  Only some of the integrated orbits
are plotted in each Figure. 

Figure \ref{fig:large}a shows test particle orbits, begun at semi-major axis
with $\Delta a/R_H = 7.5$, that experience a close encounter with the moon with 
$e=0.001$ on their first
close approach.   Close encounters occur when the moon is near
apocenter and at that time its angular rotation rate in the inertial frame  slows.  The test particles,
which are begun in circular orbits, initially have constant angular rotation rates.
In a moving and rotating frame that keeps the moon at the origin, the test particle orbits show loops.
Impacts can come from the right, and hit on the moon's trailing side, rather than from the leading side of the moon,
even though these external orbits were initially approaching the moon from the left and leading side
and outside the moon's orbit.
\citet{charnoz07} also noticed that particles
originating from outside the moon's orbit  impacted on the trailing side instead of the leading
side in simulations of an eccentric Atlas.
Figure \ref{fig:large}b is similar except $\Delta a/R_H = 2.8$ and the moon's orbital
eccentricity $e=0.0002$.
The loops are smaller in this case and one orbit passes from outside the moon to inside the moon's orbit.
Encounters such as these can facilitate the moon's radial migration (e.g., \citealt{ida00}). 

In Figure \ref{fig:eorb}a we show test particles orbits begun at 
different initial semi-major axes, with $\Delta a/R_H = 7.5$, 7.0, 6.5 or 6.0, respectively for moon 
orbital eccentricity $e=0.001$.   Figure \ref{fig:eorb}b shows orbits   with $\Delta a/R_H = 3.2$, 2.8, 2.3 or 1.8 for
orbital eccentricity $e=0.0002$.  Each panel shows test particle orbits with a single 
initial orbital semi-major axis.  The offset  $\Delta a$ in units of the Hill radius
 is labelled at the bottom of each panel.
Figure \ref{fig:eorb} shows that at each initial semi-major axis, 
impacting test particles have a range of impact longitudes on the moon, depending upon the position
of the moon in its orbit during the encounter.
We confirm the numerical results by \citet{charnoz07}, modeling Atlas' equatorial ridge, finding
that a moon on an eccentric orbit would probably accrete material covering a wider 
range of longitudes on its equator than a moon on a circular orbit.

\begin{figure*}  
\centering
\includegraphics[width=6.5in, trim={0mm 0mm 0mm 0mm},clip]{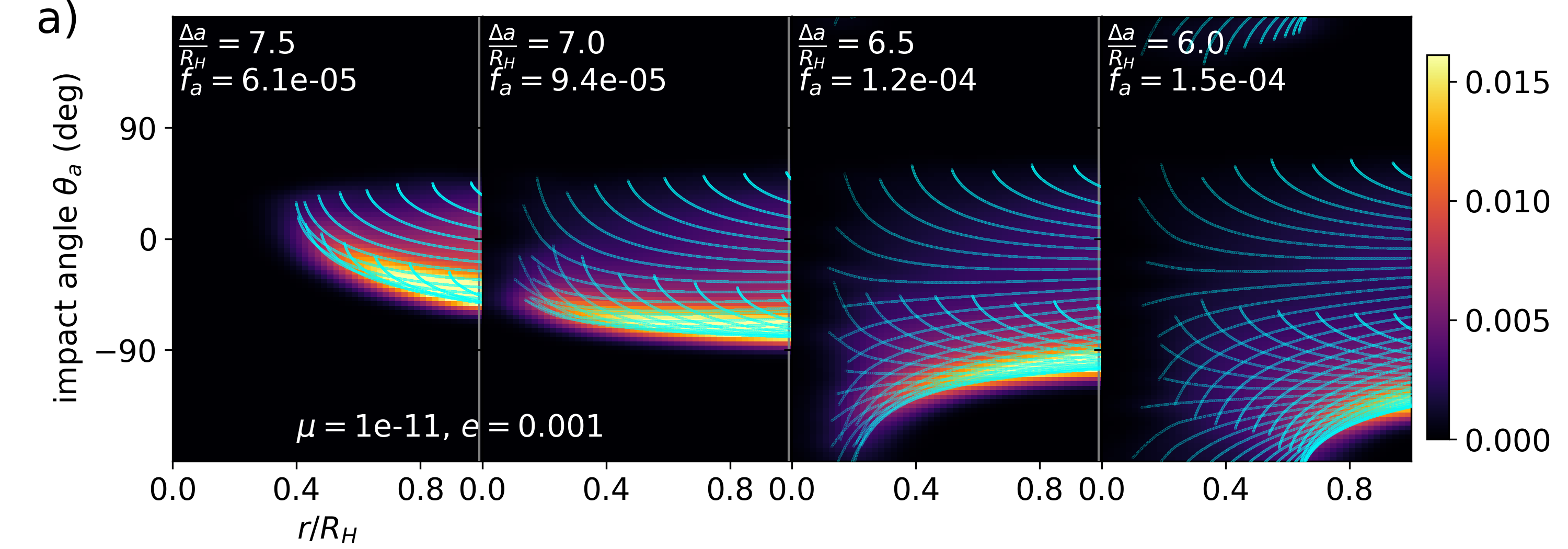} 
\includegraphics[width=6.5in, trim={0mm 0mm 0mm 0mm},clip]{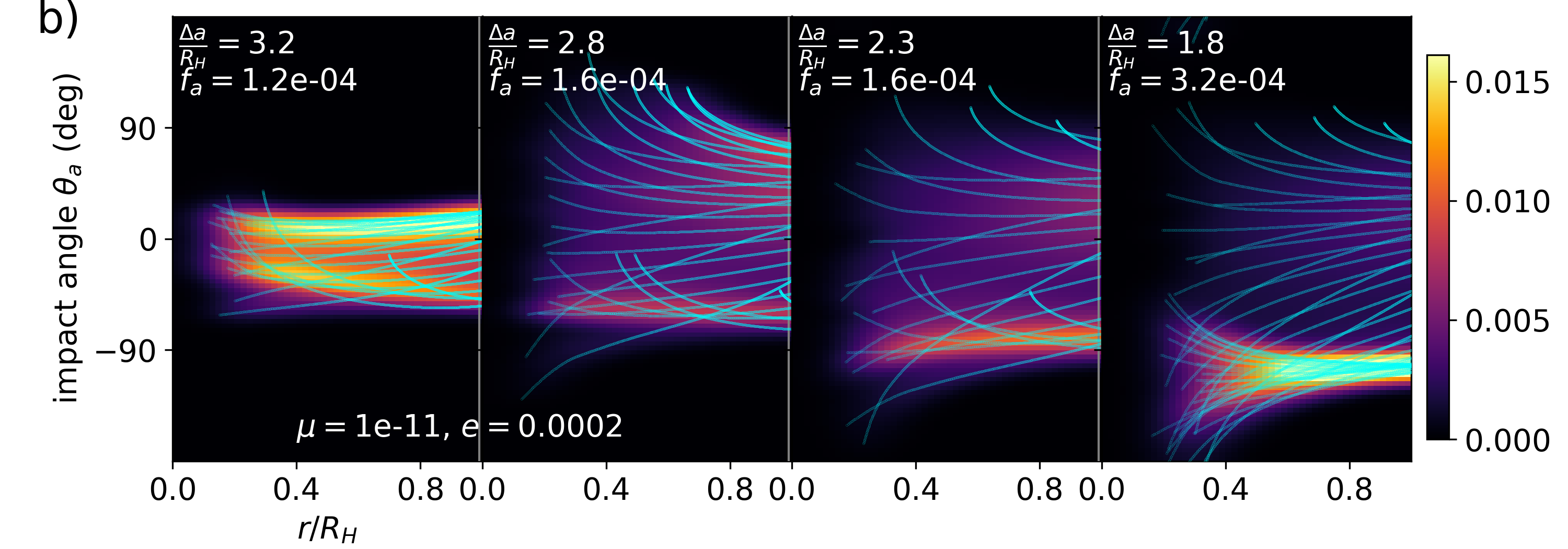} 
\caption{The impact angle distribution of material accreting onto a moon that is on 
an eccentric orbit. In each panel,  
blue-green lines show test particle orbits but as a function of impact angle ($y$-axis) 
and radius ($x$-axis) from the moon's center of mass.
The $y$ axis is in degrees and the $x$ axis in units of the Hill radius.   
The color image shows 
where mass accumulates on the surface of an accreting moon.  
The colored density images  are  histograms constructed from the number  of
orbits that crossed each radius and impact angle. 
The distribution of accreted mass on the moon's equator as a function
of the moon's longitude can be estimated
by choosing a vertical slice in the density image with radius given by the moon's surface radius.
The fraction of impacting
test particles $f_a$ is recorded on the top left of the panel.  
The colour-bar scales shows the fraction of particles that impact per degree in impact angle.
This is normalized so the integral
over impact angle of the distribution at $r=R_H$ is 1. 
The orbits are the same as shown in Figure \ref{fig:eorb} and are for a moon with mass ratio $\mu=10^{-11}$. 
a) For moon orbital eccentricity $e=0.001 \approx  7.1 e_H$. 
b) For moon orbital eccentricity $e=0.0002 \approx  1.4 e_H$.  
  \label{fig:dens}}
\end{figure*}

In Figure \ref{fig:dens} we show with blue-green lines 
the same orbits as in Figure \ref{fig:eorb} but plotted as a function
of  angle $\theta_a$ from vertical (the impact angle used in section \ref{sec:corb} and shown 
in Figure \ref{fig:pan_sil}).  The impact angle is on the $y$-axis in degrees 
and radius from the moon center is on the $x$-axis in units of Hill radius.   
These figures show the orbit impact angle as a function of radius
of the moon surface and also illustrate that for each initial orbital semi-major axis, 
impacting test particles have a range of impact longitudes on the moon's equator.

Figure \ref{fig:dens} shows that 
smaller separation in orbital semi-major axis $|\Delta a|$ gives a  lower  mean impact angle.
This is opposite to the trend seen for a moon in a circular orbit.
Since orbits were begun evenly spaced in orbital  longitude, their number density can be used to 
estimate the fraction of accreted mass that would impact the moon as a function of longitude
on its equator.
We resampled the integrated orbits  at each moon radius to produce the colored images  which show 
the number density of particles that would impact a moon  as a function of impact angle (the $y$-axis)
and for different possible moon radii (the $x$-axis).  The number density has been normalized
so that the total number density integrates to 1 at radius $r = R_H$.    Because test particles
are initially equally spaced in their circular orbits, we can compute the fraction of particles 
that impact the moon, assuming that they are initially evenly distributed over the entire circular
orbit.   The fraction of test particles $f_a$, 
that cross inside $r = R_H$ is  recorded on the top left of each figure.   
In the integrations shown in Figure \ref{fig:dens}, 
test particles on orbits initially closer to the moon are more likely to accrete onto the moon.
Equivalently, the accretion rate is higher for the lower values of offset $|\Delta a|$.

A comparison between Figures \ref{fig:eorb}a and b and between Figures
\ref{fig:dens}a and b shows that 
the trajectories at moon orbital eccentricity $e \gg e_H$ differ from those at $e \sim e_H$.
For $e=0.0002 \sim 1.4 e_H$,
the test particle loops seen in the moon's frame of reference are smaller and this gives
accretion on both leading and trailing sides, particularly at the larger values of $|\Delta a|$.     
Accretion tends to be stronger at most impact angles on the leading side for orbital eccentricity 
$e=0.001 \sim 7.1 e_H$
and $\Delta a >0$. 
At even smaller $|\Delta a|$ than shown in Figure \ref{fig:eorb} 
the most extreme point on orbit loop lies outside the Hill radius on the opposite side of the moon, so 
impacts are more evenly distributed in angle than those shown for $e=0.001$.

\begin{figure} 
\centering
\includegraphics[width=3.0in, trim={0mm 0mm 0mm 0mm},clip]{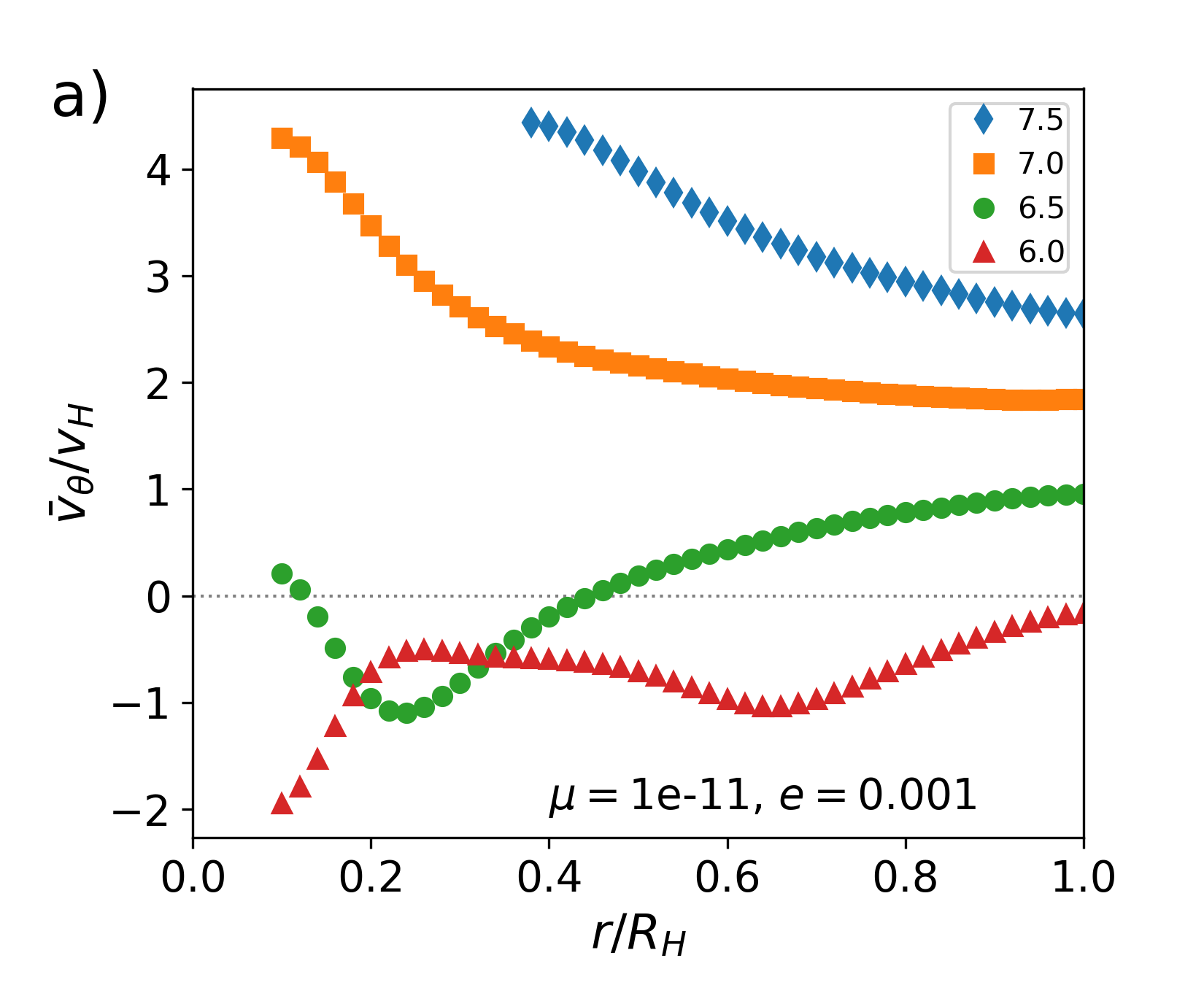}
\includegraphics[width=3.0in, trim={0mm 0mm 0mm 0mm},clip]{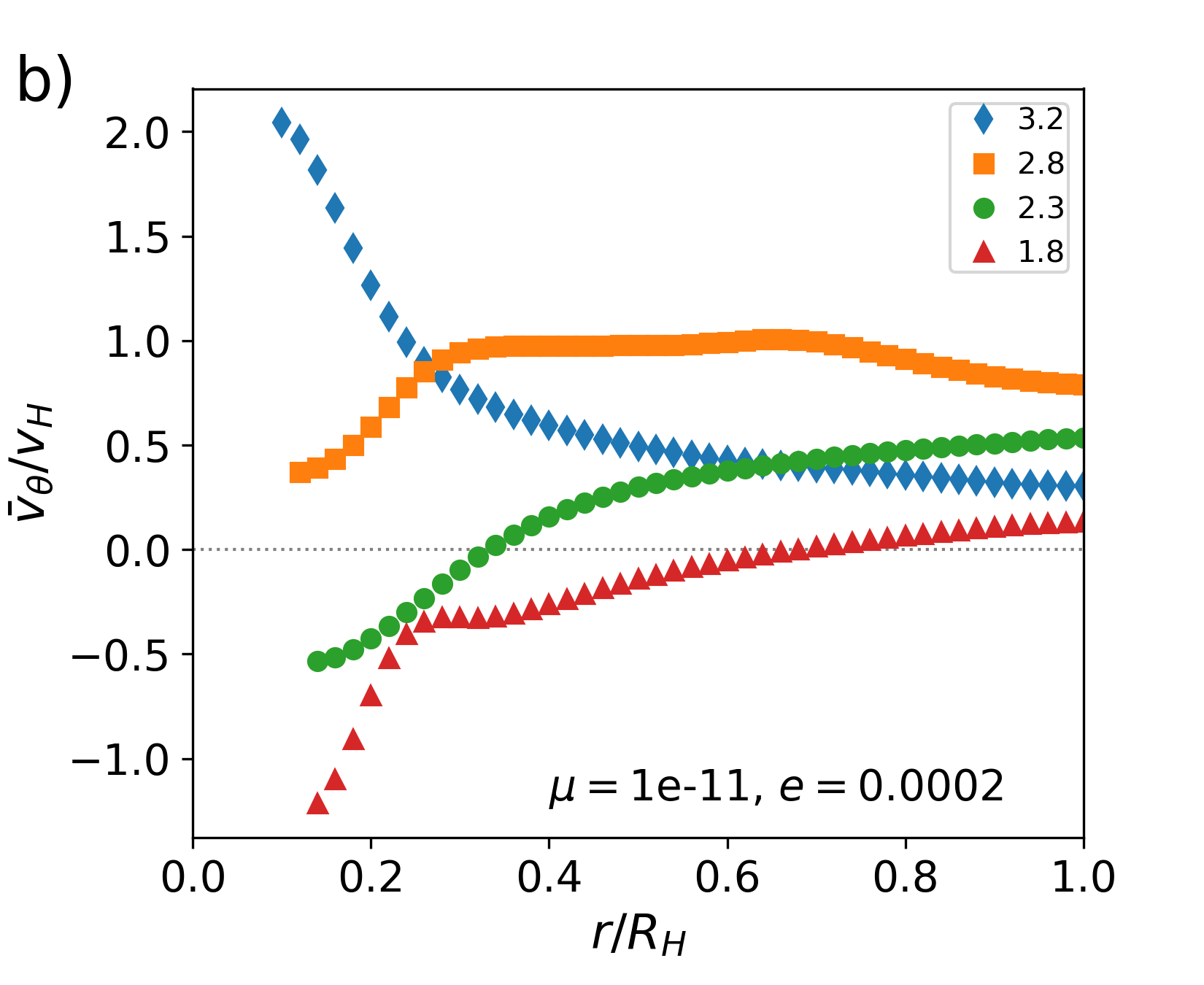}
\caption{The average value of the tangential velocity component $\bar v_\theta$ integrated 
over the accretion stream for 4 different initial test particle semi-major axis values.
Initial semi-major axes $\Delta a$ in units of $R_H$ are labelled in the keys.
The mean velocity is measured as a function of radius from the moon's center of mass
(the $x$ axis) and is in units of the Hill radius.  The mean velocity is given
in units of the Hill velocity (the $y$ axis).
The points were measured from the orbits that are also shown in Figures \ref{fig:eorb} and \ref{fig:dens}.
a) For moon orbital eccentricity $e=0.001 \approx 7.1 e_H$.
b) For moon orbital eccentricity $e=0.0002 \approx 1.3 e_H$.  
 \label{fig:vt}}
\end{figure}

As discussed in section \ref{sec:corb}, particles that impact the moon with 
positive tangential velocity component, $v_\theta >0$, could increase the moon's rotation rate.
In Figure \ref{fig:vt} we compute the average tangential velocity component, $v_\theta$,  
of the accretion stream, integrated over impact angle, and as a function of radius
from the moon's center of mass.
We show the average tangential velocity component for the different initial test particle semi-major
axis and with orbits shown in Figure \ref{fig:eorb} and for the two different moon eccentricities.
With the moon in a circular orbit, (see Figure \ref{fig:ang}b) and test particles originating from
outside the moon's orbit,
we found 
that impacting test particles  have $v_\theta >0$ for lower initial semi-major axis offset 
$|\Delta a|$.  Here with the moon in an eccentric orbit, we see the opposite
trend, $v_\theta >0$ at larger $|\Delta a|$.   This  trend is
seen because test particles tend to impact the moon at the bottom of a loop, as seen
in the frame moving with the moon, and so coming from the trailing side 
(see Figure \ref{fig:large} showing the loops).  The size of the tangential velocities 
is similar to the Hill velocity $v_H$, confirming our estimate for the critical accretion rate $\dot M_{cr}$
discussed in section \ref{sec:torque_acc}.

\begin{figure*}
\ifpreprint
\centering
\begin{tabular}{l} 
\includegraphics[width=5.0in, trim={10mm 10mm 0mm 0mm},clip]{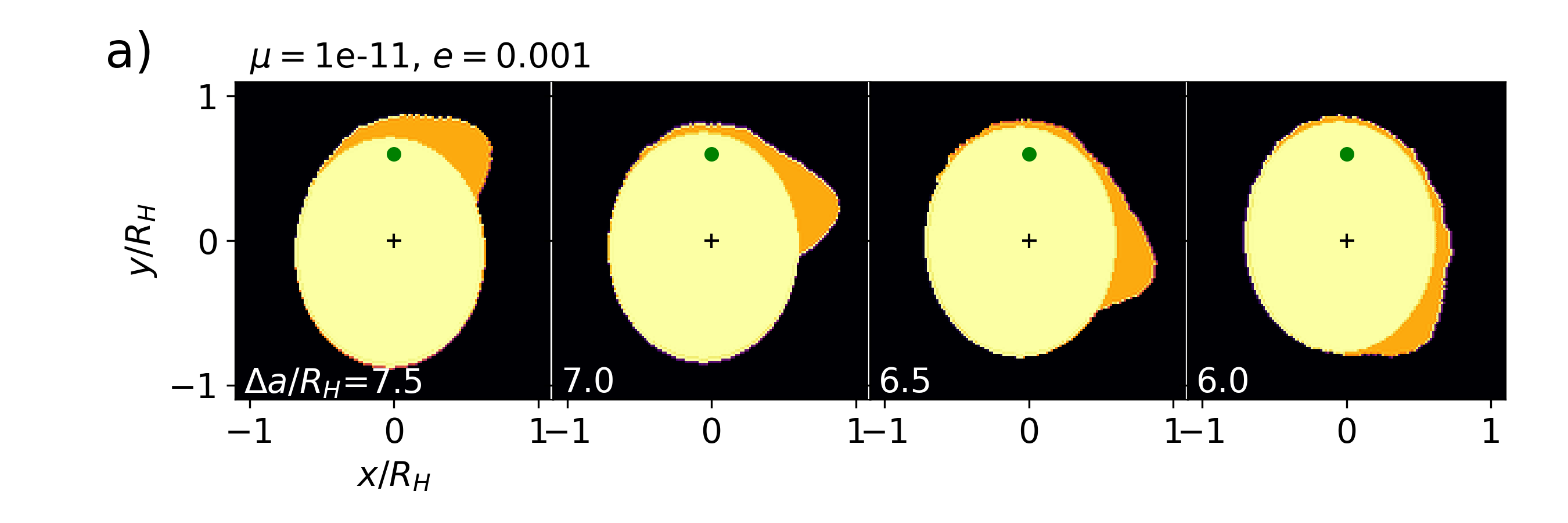} \\
\includegraphics[width=5.0in, trim={10mm 10mm 0mm 0mm},clip]{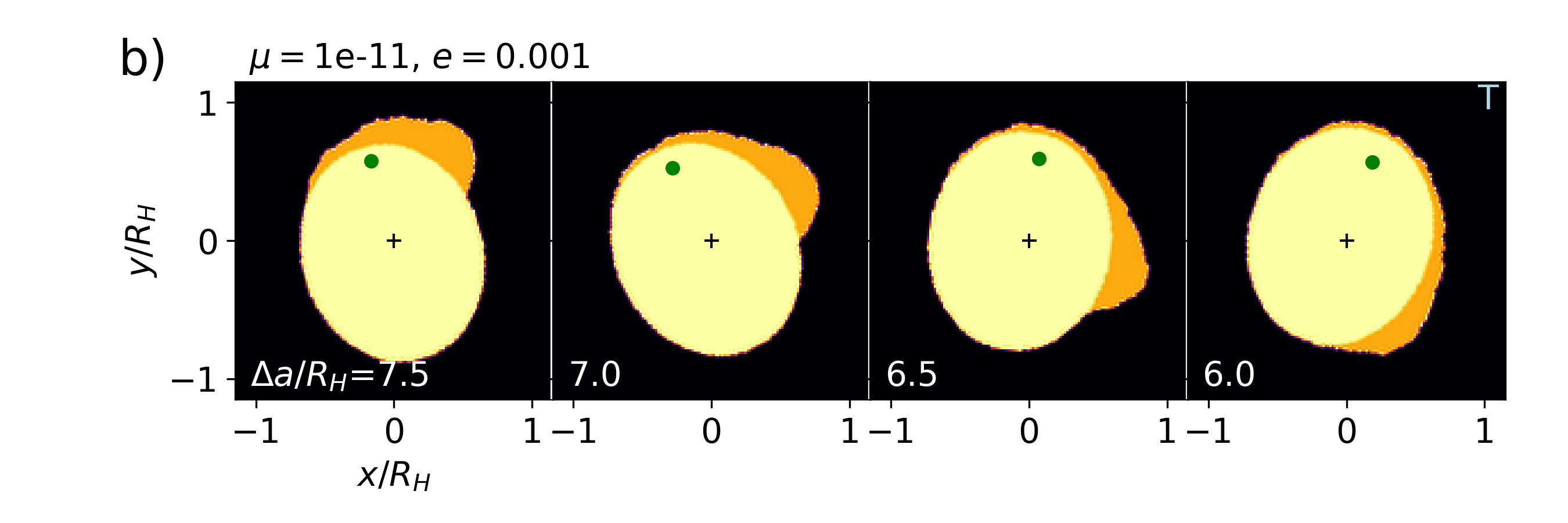} \\
\includegraphics[width=5.0in, trim={10mm 10mm 0mm 0mm},clip]{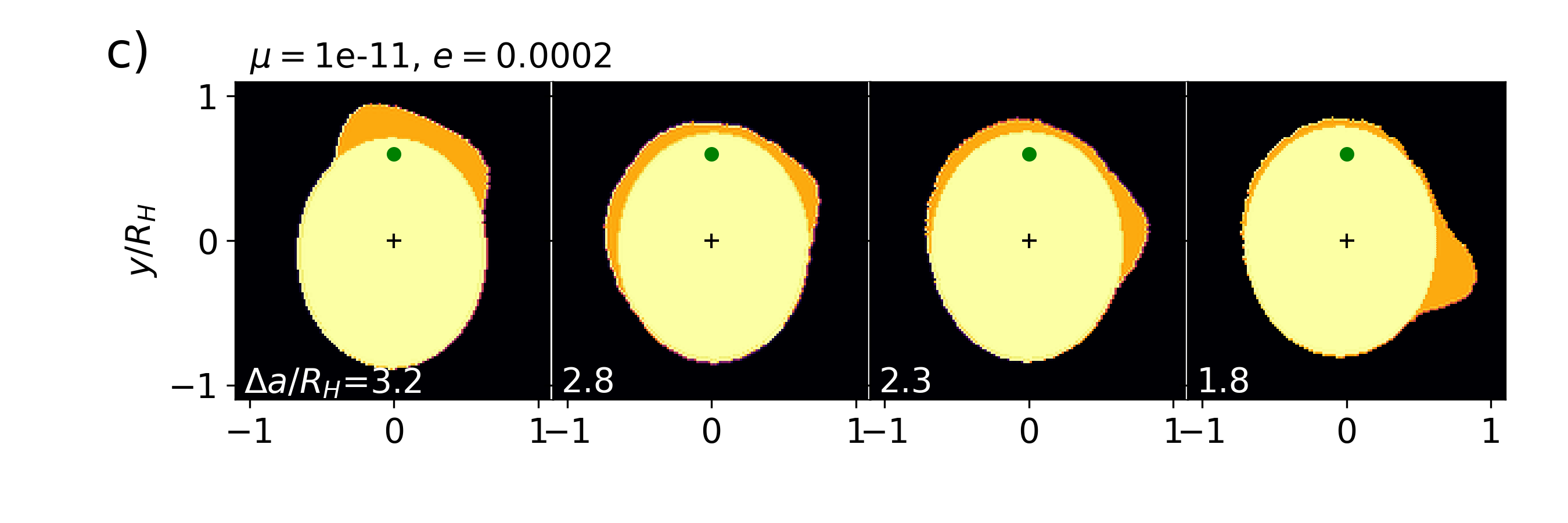} \\
\includegraphics[width=5.0in, trim={10mm  0mm 0mm 0mm},clip]{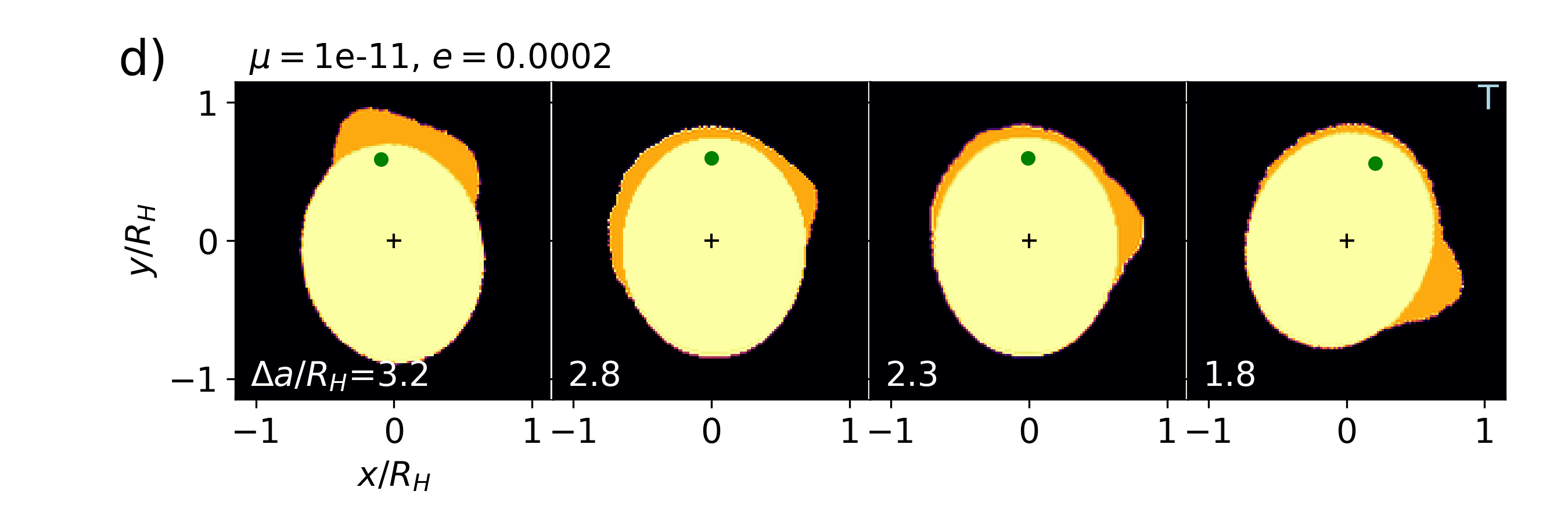} \\
\end{tabular}
\else
\centering
\begin{tabular}{l} 
\includegraphics[width=4.0in, trim={10mm 10mm 0mm 0mm},clip]{a_ephi1_notide_m.png} \\
\includegraphics[width=4.0in, trim={10mm 10mm 0mm 0mm},clip]{a_ephi1_tide_m.png} \\
\includegraphics[width=4.0in, trim={10mm 10mm 0mm 0mm},clip]{b_ephi1_notide_m.png} \\
\includegraphics[width=4.0in, trim={10mm  0mm 0mm 0mm},clip]{b_ephi1_tide_m.png} \\
\end{tabular}
\fi
\caption{The distribution of accreted mass for a moon in an eccentric 
orbit.  The moon mass ratio is $\mu=10^{-11}$.
Each row of panels 
shows material accreted from ring material initially in circular orbits 
at four different values of initial orbital semi-major
axis with initial test particle semi-major axis offset $\Delta a/R_H$ labelled on the lower left of each panel.
The total amount of accreted material is the same in all panels and is shown in orange.
The density distributions constructed from the orbits shown 
in Figure \ref{fig:dens} were used to model the longitude distribution of accreted mass.
The underlying core is shown in yellow
and has   $b_{\rm core}/a_{\rm core} = 0.83$  and $a_{\rm core} = 0.8 R_H$, based
on the shape model for Pan by \citet{thomas20}.  Axes are  in units of Hill radius.
Panels with a 'T' in the top right corners are tidally locked orientations.
a) The moon's core remains fixed during accretion and the moon's orbital eccentricity is
$e=0.001$.
b) The moon remains tidally locked during accretion and the moon's orbital eccentricity is
$e=0.001$.
c) The moon's core remains fixed during accretion and the moon's orbital eccentricity is
$e=0.0002$.
d) The moon remains tidally locked during accretion and the moon's orbital eccentricity is
$e=0.0002$.
\label{fig:ephi}}
\end{figure*}

\subsection{Accretion onto a moon that is on an eccentric orbit}
\label{sec:eacc}

\citet{thomas20} have recently updated 
shape models for Pan and Atlas.  Their new models include estimates
for ridge-removed core axis ratios which we have listed in Table \ref{tab:moon}. 
In Figure \ref{fig:ephi} we show similar elongated cores that  have  accreted mass 
along their equators.  The angular distributions of accreted mass are based  on the distributions, 
shown in Figure \ref{fig:dens}, we 
constructed from  test particle orbit integrations.  
In each panel in Figure \ref{fig:ephi}, 
the core is shown in yellow and the accreted mass is in orange. 
The green dot shows a core position that was originally on the positive $y$ axis prior to accretion.
The cores have axis ratio $b_{\rm core}/a_{\rm core} = 0.83$ and semi-major axis $a_{\rm core} \approx 0.8 R_H$
based on the ridge-removed model for the core of Pan \citep{thomas20}.   
The accretion is done in two dimensions and as described in section \ref{sec:acc_lock}.  However,
instead of allowing accretion at a single angle, 
we consecutively add mass to pixels in the mass density array using a probability that is equal to that
found at equivalent angle and radius in the density distributions that we computed from the test particle orbits.

In Figure \ref{fig:ephi}, the panels in each row show material accreted from test particle orbits at different initial orbital semi-major axes.   The initial orbital semi-major axes (described by offset $\Delta a$) 
are the same as used in 
Figures \ref{fig:eorb} -- \ref{fig:vt} and are labelled on the lower left of each panel
in units of Hill radius.  In Figure \ref{fig:ephi}a,c the body core remains at a fixed angle with respect
to the moon/planet line which is the vertical direction.   
In Figure \ref{fig:ephi}b,d the entire body remains tidally aligned, keeping 
principal body axis aligned with the moon/planet line.  
Figures \ref{fig:ephi}a,b are for a moon with orbital eccentricity $e=0.001\approx 7.1 e_H$ and 
Figures \ref{fig:ephi}c,d are for a moon with orbital eccentricity $e=0.0002\approx 1.4 e_H$. 

As described in section \ref{sec:acc_lock},
the center of mass position is computed after each parcel of mass is added.  
For  Figure \ref{fig:ephi}b and d, body principal axes are recomputed between
each body accretion event, so as to maintain tidal alignment.
We have neglected the change in the tidal orientation direction in the moon's orbit known as the
librational tide since the orientation angle change would be approximately the same size 
as the orbital eccentricity and negligible.

The distributions of accreted material in Figure \ref{fig:ephi}a and Figure \ref{fig:ephi}b are similar. 
Pan's core is sufficiently elongated that the accreted mass
does not cause a large shift in the body orientation. 
The mass in Pan's lobes seem to increase in with increasing western longitude,
and the accreted wedges in Figures \ref{fig:ephi}a,b exhibit the same
trend.   This suggests that the shape of the lobes on Pan's equatorial ridge
is related to the angular distribution of impacting
material.  At lower orbital eccentricity,  Figures \ref{fig:ephi}a,b show more complex morphology
that we attribute to the two lobes of the accretion stream that are seen
in Figure \ref{fig:eorb}b and \ref{fig:dens}b, particularly at larger $|\Delta a|$.


In this section we have looked at test particles in initially circular orbits originating outside the moon's orbit  
 that impact  a moon in an eccentric orbit. 
 With test particles originating inside the moon's orbit, similar phenomena is observed after rotation by $180^\circ$ about the moon center of mass.   
We have also integrated test particle orbits that were initially on eccentric orbits that impact a
moon that is in a circular orbit.    Similar orbit loops are observed for these if the same
eccentricity is used for the test particle orbits  when doing the comparison.  This similarity
was also reported by \citet{weiss09}.  
In other words, we compared orbits of 
 test particles with  orbital eccentricity $0.001$ impacting a moon on a circular orbit to
orbits of test particles on circular orbits impacting a moon with  orbital eccentricity
$e=0.001$.   

In summary, for a moon in eccentric orbit with eccentricity $e>e_H$ or $e \sim e_H$, test
particles in initially circular orbits have a different distribution of impact angles 
than those encountering a moon in a circular orbit. 
 Test particles at the same initial semi-major axis, but
different initial orbital longitudes impact the moon at different longitudes on the moon's 
equator.   For moon orbital eccentricity $e> e_H$, 
particles initially exterior to the moon's orbit tend to impact the moon on the trailing side. 
The angular distribution
of impacts peaks nearer the sub-Saturn point for lower initial semi-major axis, $\Delta a$
external to the moon's orbit. 
For $e \sim e_H$, the accretion stream can have two lobes, one impacting
on the trailing side and the other on the leading side.
Pan's equatorial ridge lobes seem to increase in radius with increasing western longitude,
and accreted wedges constructed using the angular distribution computed from 
 impacting orbits originating at a single initial orbital semi-major axis
 exhibit the same trend for $e>e_H$, but can have more
 complex shapes if $e \sim e_H$.  

\begin{figure*} 
\centering
\begin{tabular}{ll} 
\includegraphics[width=3.0in, trim={0mm 0mm 0mm 0mm},clip]{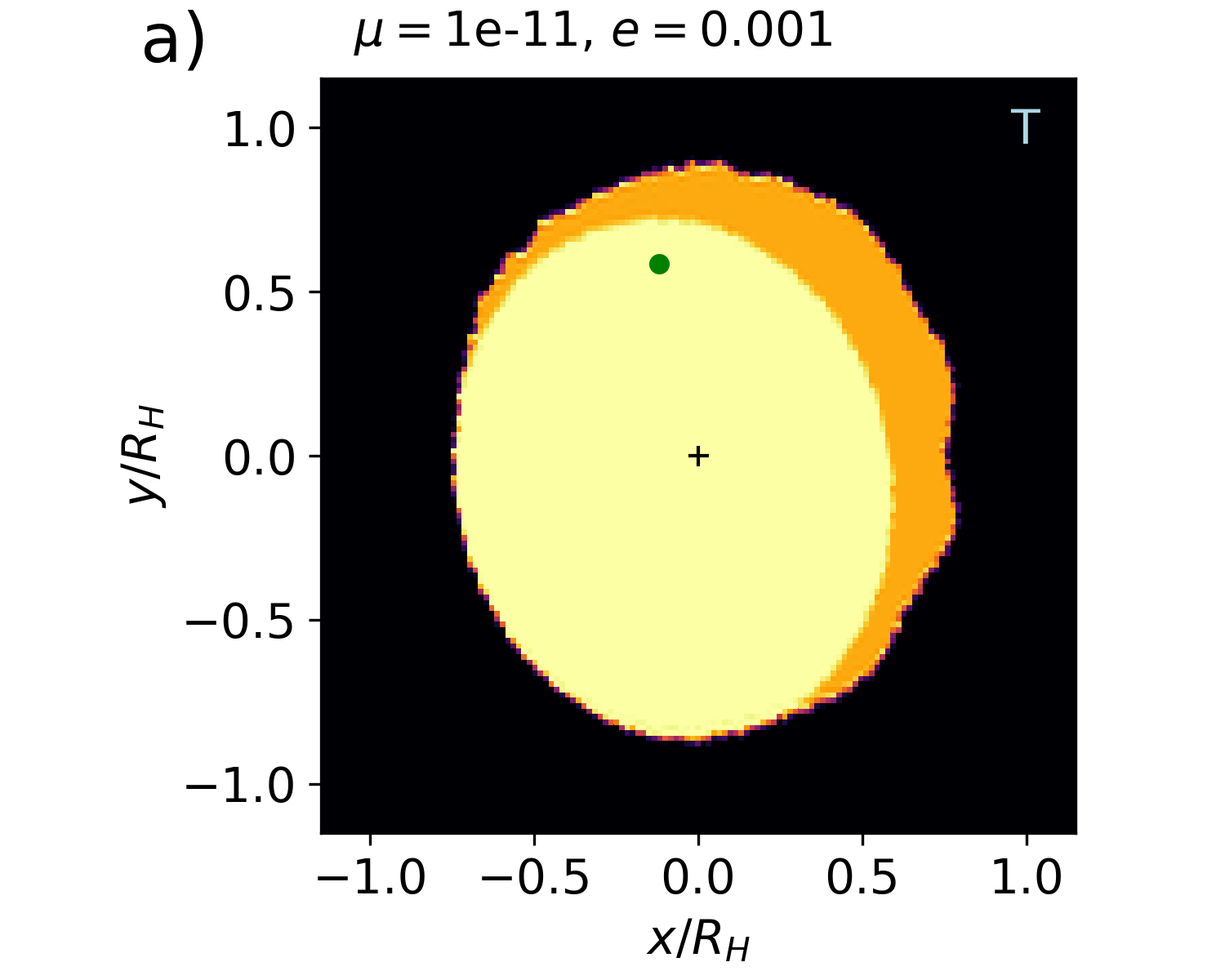} &
\includegraphics[width=3.0in, trim={0mm 0mm 0mm 0mm},clip]{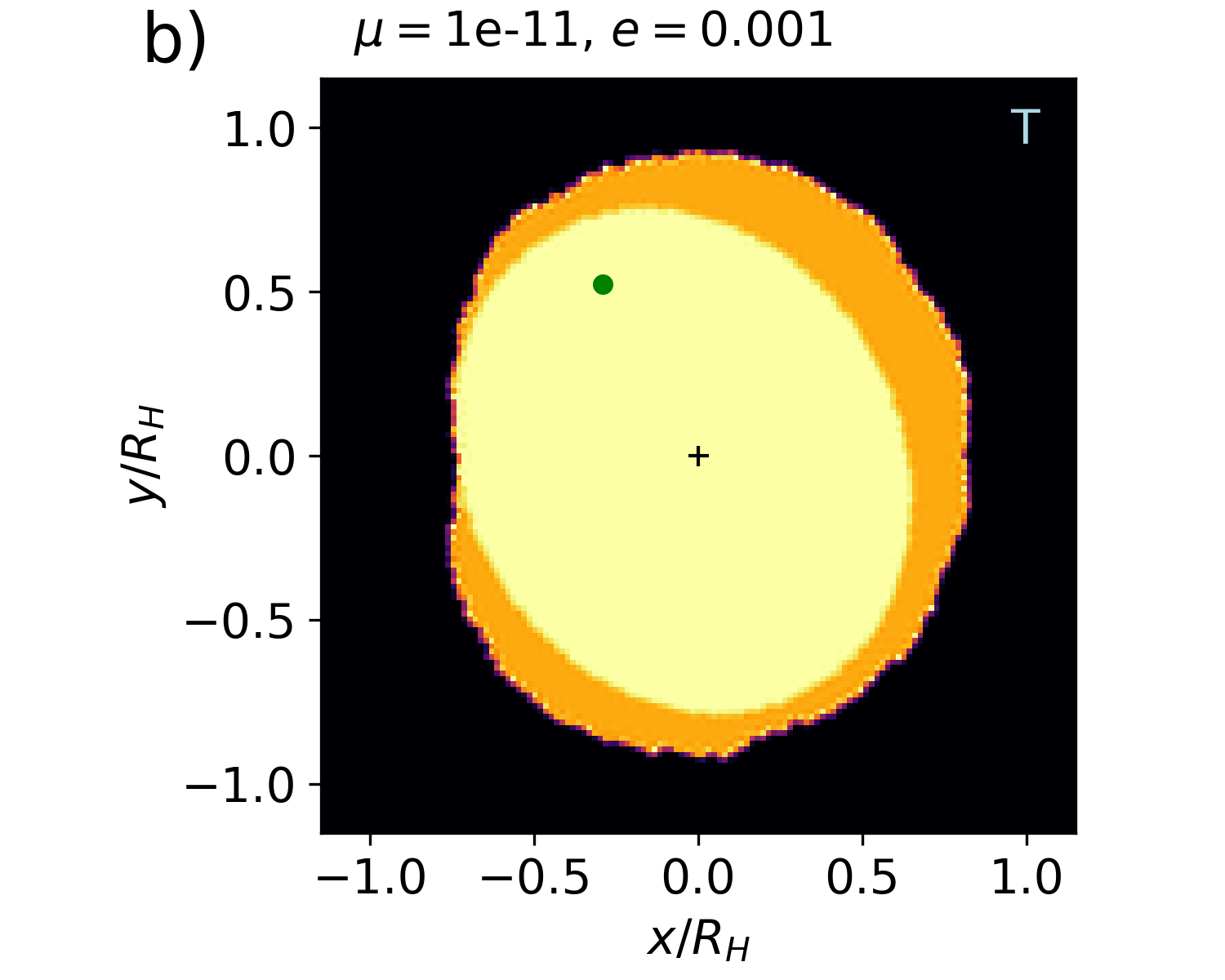} \\
\includegraphics[width=3.0in, trim={0mm 0mm 0mm 0mm},clip]{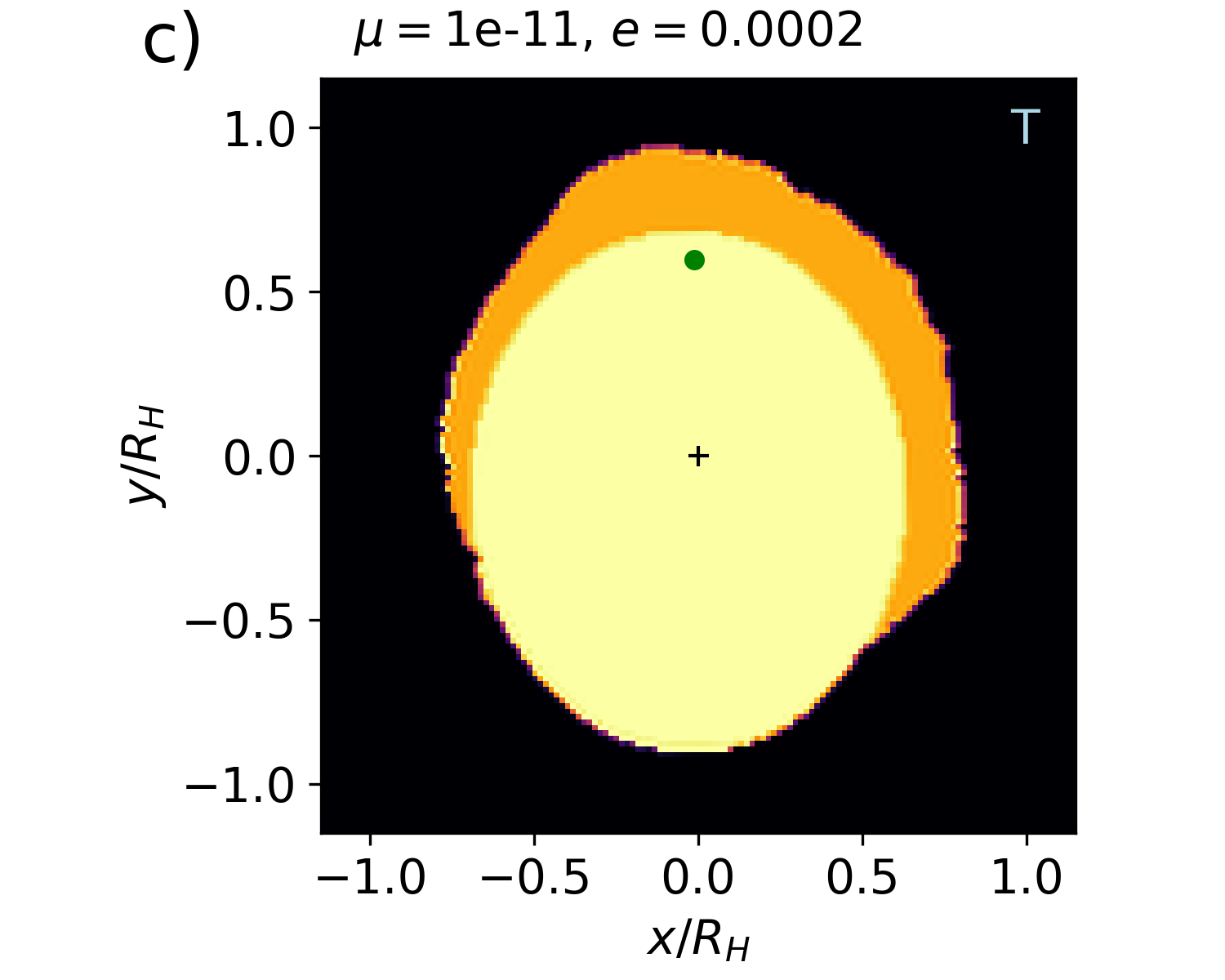} &
\includegraphics[width=3.0in, trim={0mm  0mm 0mm 0mm},clip]{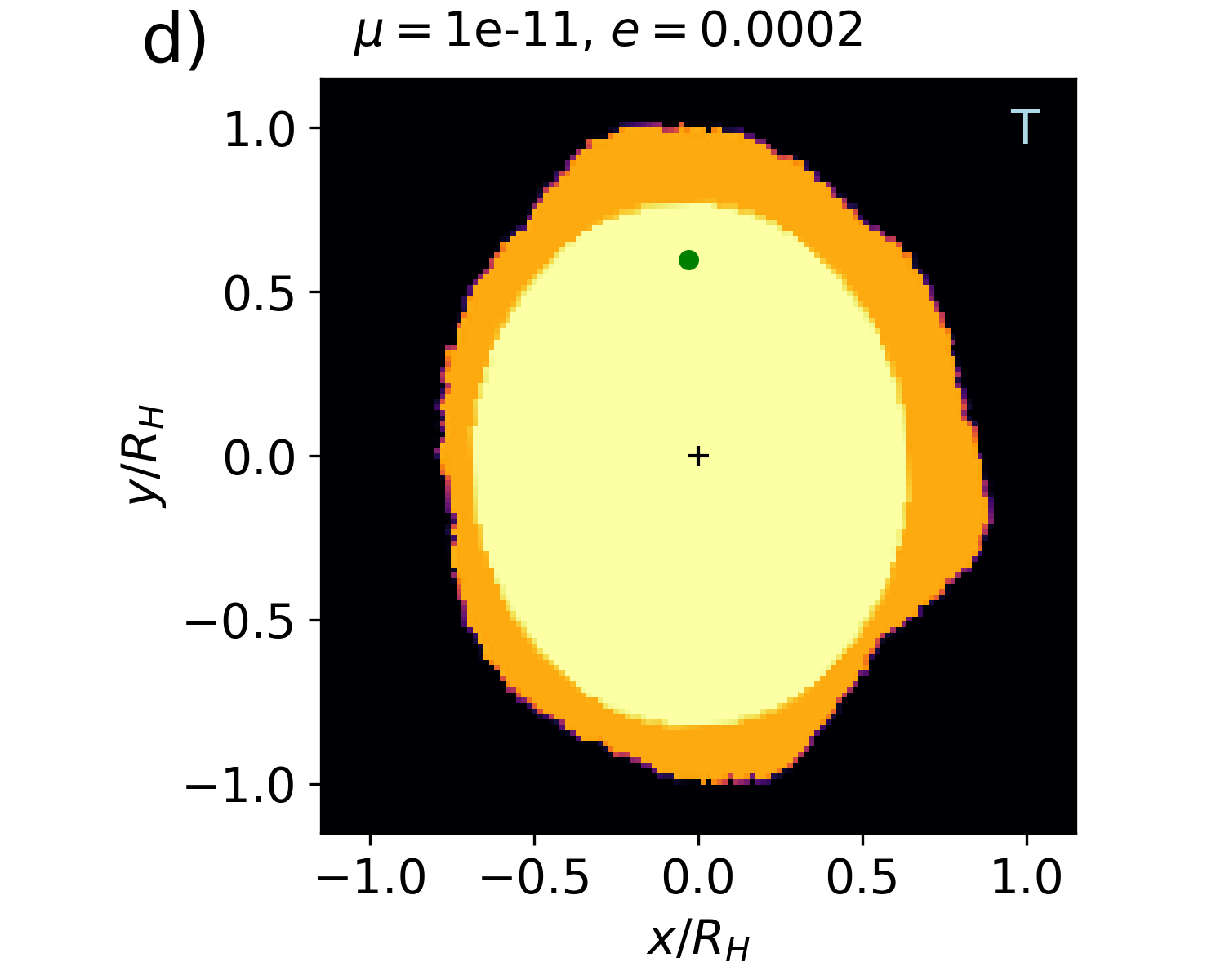} \\
\end{tabular}
\caption{The distribution of accreted mass for a moon in an eccentric 
orbit and with roughly simulated gap profiles. The moon mass ratio is $\mu=10^{-11}$.
The total amount of accreted material is the same in all panels and is shown in orange.
The density distributions constructed from the orbits and shown 
in Figure \ref{fig:dens} were used to model the longitude distribution of accreted mass.
The underlying core is shown in yellow
and has   $b_{\rm core}/a_{\rm core} = 0.83$  and $a_{\rm core} = 0.8 R_H$.
Axes are  in units of Hill radius.
In all cases the accreting moon remains oriented as if it were tidally locked, denoted with a 'T' on the top right corners of the panels.
a) The moon only accretes from its outer disk edge and the moon's orbital eccentricity is
$e=0.001$.
b) The moon accretes from its outer disk edge and with a wider gap from
its inner disk edge.  The moon's orbital eccentricity is
$e=0.001$.
c) The moon only accretes from its outer disk edge and the moon's orbital eccentricity is
$e=0.0002$.
d) The moon  accretes from its outer disk edge and with a wider gap from
its inner disk edge.  The moon's orbital eccentricity is
$e=0.0002$.
\label{fig:dtry7}}
\end{figure*}

\subsection{Accreting from a gap edge}
\label{sec:e_edge}

In the previous section we examined accretion from  a  ring of disk 
material that originates from a single initial orbital semi-major axis.  We now consider
accretion from a multiple orbital semi-major axes, mimicking accretion
from a gap edge that would be described by a mass surface density profile $\Sigma(\Delta a)$.

To mimic accretion from an outer edge of a gap, the accreted mass distributions in Figure \ref{fig:ephi}a or c are combined after they have been multiplied by $\Sigma(\Delta a) f_a$ to take into 
account the gap edge mass surface density profile and the fraction of particles $f_a$, at each initial orbital semi-major axis, that impact the moon.
The fractions of impacting particles are labelled on Figure \ref{fig:dens}.
For moon eccentricity $e=0.0002$, the fraction of particles that impact is two to three times higher   
at semi-major axis offset $\Delta a/R_H = 1.8$ than at the higher values (2.3, 2.8 and 3.2).
For moon eccentricity $e=0.001$, the fraction of particles that impact
is about twice as high at $\Delta a/R_H = 6.0$ than at 7.5.
The impact fractions increase with decreasing $\Delta a$,  opposite to
the drop expected in the density  $\Sigma(\Delta a)$ in an outer gap edge.
This would decrease the contrast between summed distributions. 

In Figure \ref{fig:dtry7} we show accreted mass distributions 
 constructed from the density distributions shown 
 in Figure  \ref{fig:dens} and \ref{fig:ephi}.  Each panel uses distributions
 from the four different initial test particle
semi-major axes also shown in these figures. 
The equatorial ridges were accreted onto an underlying core with 
 $b_{\rm core}/a_{\rm core} = 0.83$  and $a_{\rm core} = 0.8 R_H$, as previously used in Figure \ref{fig:ephi}.
While accreting, the body principal axis remains aligned vertically, to mimic tidal lock.
For Figures \ref{fig:dtry7} a,b, we use the density distributions for moon
eccentricity $e=0.001 \approx 7.1 e_H$
and for Figures \ref{fig:dtry7} c,d we use those for $e=0.0002 \approx 1.4 e_H$.
To mimic accretion from the inner edge of a gap, we use
the density distributions constructed for the outer edge and rotate the accretion stream
and angular distributions 
by $180^\circ$.

In  Figure  \ref{fig:dtry7} a,c the moon accretes only from an outer gap edge and with
$\Sigma(\Delta a) f_a$ assumed to be constant at each of the four $\Delta a$ values
used to compute the angular distribution of accreting material.
In Figure  \ref{fig:dtry7} b,d, in addition to accretion from an outer
gap edge, accretion is also from an inner gap edge, but the distance to
the inner edge  is wider
and accretion is allowed only from the outermost two semi-major axes present  
in Figure \ref{fig:dens}a or b. 

For moon orbital eccentricity $e > e_H$, accretion from gap edges gives
a fairly smooth equatorial ridge.   If the inner gap edge is more distant
from the moon and the outer gap edge profile is flatter than the inner gap edge profile, 
(Figure \ref{fig:dtry7}b) then the equatorial ridge has  a depression on the leading
side, like that of Atlas.   \citet{charnoz07} suggested that the leading/trailing asymmetry 
of Atlas's equatorial ridge was due to its orbital eccentricity.  
However, we and \citet{weiss09} find that near a moon on an eccentric orbit,  test particle orbits 
originating interior to a gap resemble those originating exterior to a gap after rotation by $180^\circ$.   
The leading/trailing symmetry of Pan's equatorial ridge
could be due to an asymmetry in the outer and inner gap edge density profiles.
We found that the leading side would accrete less for a moon
in a circular orbit if the outer gap edge were more distant from the moon's orbit (see
section \ref{sec:c_edge}).    
With moon eccentricity $e>e_H$,
we find the opposite is required to match Atlas, 
the inner gap edge must be more distant from the moon.

Figure \ref{fig:ephi}c  is  interesting as a  sum of the different distributions  
in each panel would not produce an even longitude distribution.
 The  $\Delta a/R_H=1.8 $ panel (the rightmost panel in Figure \ref{fig:ephi}c)  
 shows accreted mass concentrated at a longitude of approximately 300$^\circ$ W (on 
the lower right) and near where Pan shows a lobe in its equatorial ridge on its trailing side.
The middle two panels (with $\Delta a/R_H=2.8 $ and 2.3) 
have wider  longitude distributions of accreted material.
If the mass surface density profile of the gap's outer edge extends to $\Delta a/R_H = 1.8$
then the equatorial ridge might gain a lobe at about 300$^\circ$ W.
When we 
combine the angular distributions at  4 different semi-major axes
and for moon eccentricity $e = 0.0002$  (see Figures \ref{fig:dtry7}c,d), the equatorial ridge has
more structure than at higher eccentricity  (Figures \ref{fig:dtry7}a,b).
Figure \ref{fig:dtry7}d allowing accretion from both outer and inner gap edges
but with a wider inner gap, exhibits   multiple lobes with peaks, like on Pan.
We see more lobate structure at $e \sim e_H$ than at $e>e_H$. This is likely 
because loops in the orbit (as seen in the moon's frame) 
are about the same size as the Hill radius when $e \sim e_H$ whereas they
are larger than the Hill radius when $e >e_H$.
The ridge in Figure \ref{fig:dtry7}d 
is perhaps similar to the ravioli shape of Pan's equatorial ridge.  
During equatorial ridge accretion, the gap profile and orbital eccentricity need not be fixed.
With a varying gap profile and eccentricity, more complex shaped equatorial ridges might be accreted.
 
In summary, we find that an equatorial ridge with 
complex morphology  
might be accreted from ring material located in the
edges of a gap, if the moon has an  
 orbital eccentricity.    A smooth equatorial ridge with a depression
 on the leading side, might arise if accretion took place at moon orbital
 eccentricity $e>e_H$ and a gap that is wider inside the moon's orbit
 than outside it.  A multiple lobed equatorial ridge, resembling
 that of Pan, could be accreted at $e\sim e_H$, also
 with  an asymmetric  gap surface density profile.


\section{Scenarios for accreting equatorial ridges}
\label{sec:scen}

A number of possible scenarios are suggested by the illustrations we
have constructed showing the distribution of accreting material estimated from
impacting orbits.
With both moon and ring material on  circular orbits, and a fixed
distance between ring edge and moon semi-major axis, a multiple lobed  equatorial 
ridge, like Pan's,  probably cannot be accreted onto a tidally locked moon.    
We consider the following scenarios that relax the aforementioned restrictions:
\begin{itemize}
\item The moon's orbital eccentricity  is increased due
to an impact, pushing it into ring material.
\item The moon is captured into a spin-orbit resonance and accretes while in this resonance.
\item The moon's eccentricity is increased due to orbital resonance with another moon
and its eccentricity pushes it into the ring so that it can accrete.
\item The moon migrates through the ring.  The migration itself maintains
a narrow gap between ring edge and moon, facilitating accretion.
\end{itemize}

\subsection{An impact causes an increase in orbital eccentricity}
\label{sec:impact}

We first consider an impact mediated scenario.
An impact could vary both moon orbital eccentricity and spin simultaneously; however, the eccentricity must be maintained over about $10^5$ years for the accretion to take place slowly enough  to satisfy the constraint on the accretion rate ensuring tidal lock, $\dot M < \dot M_{cr}$,  from equation \ref{eqn:Mcrit}.   Recall that if the moon starts spinning while accreting, the accreted
material would be more evenly distributed and would probably lack lobes, peaks, divots or other longitudinal substructures.  As discussed in section \ref{sec:edamp}, 
the estimated eccentricity damping time due to excitation of spiral density waves is  
only a few thousand years, so is shorter than the required accretion time, 
except possibly for Atlas which 
is in a lower ring density region outside the A-ring.  
The time to regain tidal lock is $10^5$ to  $10^6$ years (see Table \ref{tab:computed}), 
so the orbital eccentricity would be damped before the moon regains tidal lock.
If orbital eccentricity is required to give the moon a source for ring material to accrete, 
then its rapid damping presents a problem for an impact triggered scenario.

\subsection{Spin-orbit resonance}
\label{sec:spinorbit}

We consider a scenario where the moon spends time in a spin-orbit resonance. 
If Pan were captured into a spin-orbit resonance, instead of accreting along 
a single angle, material might impact at multiple angles on the body. 
Pan's current angular rotation rate is about 0.38 times the centrifugal break-up spin 
value for an equivalent density sphere,  so Pan cannot spin more than three times faster 
than its mean motion.  It could enter the 1:2 spin orbit resonance where the body rotates
once every two orbits.
To enter a spin-orbit resonance, the moon must first be pulled out of tidal lock.  
This could happen because of an impact or a high enough accretion rate.
Once a moon is pulled out of tidal lock, it must tumble prior 
to reaching the spin synchronous state \citep{wisdom87} or being captured 
into a spin-orbit resonance.  
%

  
To mimic accretion within spin-orbit resonance,  we can consider a  
model where accretion takes place at impact angle $\theta_a$ but is consecutively rotated 
by an integer fraction of $2 \pi$ between each mass accretion event. 
Because accretion occurs at symmetrical points around the equator, the orientation of the body principal axes should not change significantly during accretion.    
If there is structure in the accreted lobes, they must be present in the accretion stream 
as body orientation  (taking into account 
the periodicity of the resonance)  is unlikely to shift significantly during accretion. 

The most recent shape model for Pan has peaks at longitudes near  0$^\circ$ and 180$^\circ$ W, on the   Saturn-facing and anti-Saturn facing sides.   Two additional  peaks are found on the trailing side at longitudes of about 240$^\circ$ and 300$^\circ$W. The trailing side suggests the moon equatorial ridge is a six-peaked polygon. However, the most recent shape model \citet{thomas20} (shown in silhouette in   Figure \ref{fig:pan_sil})  exhibits only a single peak on the leading side  at about 100$^\circ$W, instead of two peaks.  This lack of symmetry implies  that a spin-orbit resonance is not a good explanation for Pan's polygon-shaped equatorial ridge. 
Scenarios for accreting complex equatorial ridges involving spin-orbit resonance for Pan's complex
equatorial ridge seem to be excluded. 

\subsection{Accretion due to excitation of moon orbital eccentricity by orbital resonance} 
\label{sec:ecc}

We consider scenarios where a moon's eccentricity
is excited due to orbital resonance with another moon.  With an increase in orbital
eccentricity, the moon could graze  the edges of its hosting gap in the  rings 
at orbital apocenters and pericenters, where it could accrete ring material.
Epochs of moon migration, even on the slow viscous migration timescale,    
could push moons into pairwise orbital resonances. 
Migration can occur in either direction,  so a  moon's orbital
semi-major axis can approach that of 
another moon leading to resonance capture and causing an increase
in orbital eccentricity. While captured
in a mean motion resonance,  orbital eccentricity can be maintained or  increased
if migration continues.  This mitigates the short eccentricity
damping timescale discussed in section \ref{sec:edamp}. 
The outer region of the A-ring, in which Pan and Daphnis reside, 
is particularly rich in first-order Lindblad resonances with Pan, Atlas, Prometheus, Pandora, Janus, Epimetheus, and Mimas \citep{tajeddine17b}, and Atlas's eccentricity is currently 
enhanced due to orbital resonance with Prometheus \citep{cooper15}.
Small movements in the moon semi-major axes could bring two bodies into resonance.


While a moon's orbital eccentricity can be maintained by orbital resonance, in the absence
of migration, a gap hosting the moon would widen. Would the gap become
sufficiently wide that accretion ceases, or can accretion be maintained long enough
to accrete an equatorial ridge?
A moon embedded in a ring causes radial variations on the edges of its gap 
\citep{showalter91,porco05,porco07,weiss09}.
A close encounter with the moon causes a change in a ring particle's
semi-major axis $\Delta X$ that has been used to estimate
the torque on the ring material that pushes the ring away from
the moon \citep{goldreich82,weiss09}.

If the moon is on an eccentric orbit then the edge particle's shift $\Delta X$ depends
on the phase of the encounter \citep{weiss09}.  The total torque exerted
on the disk material, integrated over ring longitude,
 may not be strongly dependent on the moon's eccentricity \citep{sari03}.
However the orbit integrations by \cite{weiss09} suggest
that locations or distances from the moon's semi-major axis where the torque is exerted 
onto the ring material could 
depend on  eccentricity.    The torque from the moon
onto the ring is probably exerted onto the disk over a larger
range in semi-major axis offset $\Delta a$ if the moon is on an eccentric orbit, than if the
moon is on a circular orbit.
Unfortunately, predictions of gap surface density profiles
that balance the torque caused by the moon
onto the ring against viscous diffusion and accretion 
usually assume the moon is in a circular orbit \citep{gratz19b}.  This is usually
a good assumption because
of the short eccentricity damping time (as we discussed in section \ref{sec:edamp}). 
So far as we know, gap edge surface density profiles for gaps
opened by eccentric moons have not been predicted analytically. 

Moons  are characterized as able to open a gap or not, depending
upon the moon mass ratio and viscosity in the disk \citep{lissauer81}.
An eccentricity of a few $e_H$ would let Pan or Daphnis 
graze their  gap edges.  If the torque exerted on the ring material is
not a strong function of orbital eccentricity and the eccentricity is maintained due
to orbital resonance with another moon, then the gaps opened by Pan or Daphnis would
not be wide enough to stop accretion.  As long as the moon eccentricity can be maintained,
accretion via grazing gap edges would seem to be a viable scenario for accretion
of an ornamental equatorial ridge.


\subsection{Accreting while undergoing radial migration}
\label{sec:mig}

We  consider equatorial ridge accretion scenarios involving a migrating moon. 
An advantage of a migration scenario is that proximity of dense and cold
ring material to the moon can be prolonged by the migration itself.  With little or no 
migration, ring material is pushed
away from the moon's orbit and its hosting gap widens due to scattering and  
torques associated with spiral density waves that are driven by the moon itself.
The gap widens until spiral density wave driven torques
balance those associated with viscous spreading and an equilibrium state is achieved. 
The Keeler and Encke gaps are wide enough to prevent accretion onto
Daphnis and Pan, though Pan's gap is not entirely empty \citep{hedman13}.
If the moon migrates quickly, it can remain in proximity to ring
material or only open a narrow gap (see Figure 4 by \citealt{bromley13} showing
an N-body simulation of a moon rapidly migrating into a particulate disk).
The gap of a migration moon can be asymmetric and contain substructures
such as feathers \citep{weiss09} or streams (Figure 4 by \citealt{bromley13}).
In section \ref{sec:c_edge} and \ref{sec:e_edge}, we found that accretion from a 
gap  that has 
asymmetric surface density edge profiles might account for Pan and
Atlas' equatorial ridge morphology.  Migration might
naturally cause a difference between the density profile of the gap edge that the moon
is migrating toward compared to the one on the  opposite side. 

Pan, Daphnis and Atlas are too massive to migrate via the type I migration
mechanism where the body remains embedded within the ring material \citep{bromley13}.
If they opened a gap and migrated via a type II mechanism and at the viscous
evolution rate (about $10^{-6}$ km/yr, eqn 28 by \citealt{bromley13}), 
 then little ring material would have been near enough to the moon to accrete onto it.
The type III migration mechanism \citep{masset03} requires ring material to pass near the moon 
\citep{ida00,masset03},  so accretion  onto the moon could
be a natural consequence of this mechanism.
\citet{bromley13} estimate that both Pan and Daphnis could undergo 
rapid migration, at a semi-major axis drift rate of about 
$\frac{da_{\rm mig}}{dt} \sim 20$ km/yr, and in the mode of
type III migration where corotating material is gravitationally pulled from one
side of the orbit to the other \citep{ida00,masset03}.
Estimates of the  type III migration rate are based on the moon/particle encounter frequency in the moon's corotation zone and so proportional to the ring mass surface density, $\Sigma_{\rm ring}$
  \citep{ida00,bromley13}.  
  
In section \ref{sec:torque_acc} we estimated that the duration of equatorial ridge
accretion was at least $10^5$ years.
An issue with a rapid migration rate of 20 km/yr is that in $10^5$ yr, 
the moon could pass entirely through Saturn's rings.  However, 
  type III radial migration can go in either direction, inward or outward, 
  depending upon how it is triggered \citep{bromley13}. 
Type III migration requires a cold particulate disk and  is prevented or slowed by inclination and  eccentricity excitation of particles in the ring. 
As the A-ring exhibits variations in density  with abrupt transitions or jumps
at the locations of gaps and resonances  \citep{tajeddine17b,tiscareno18},   a migrating  moon would likely experience variations in migration rate. 
If migration can be slower at times and with 
 reversals, then perhaps Atlas, Pan and Daphnis 
 could have experienced episodic events of fast migration.

How much ring mass passes through the moon's orbit per unit time during type III migration?
We can estimate this rate from the migration rate and ring surface density 
\begin{align}
 \dot M_{\rm cross} &= 2 \pi a \frac{da_{\rm mig}}{dt} \Sigma_{\rm ring} \nonumber \\
  &= 3 \times 10^{14} \ {\rm kg/yr} 
  \left( \frac{a}{a_{Pan}}\right)
   \left( \frac{da_{\rm mig}/dt}{1 \ {\rm km/yr}}\right)
   \left( \frac{\Sigma_{\rm ring}}{ 400 \ {\rm kg\ m}^{-2}}\right)    . \label{eqn:Mor}
\end{align}
The accretion rate onto the moon would be a fraction of this mass rate, $\dot M = f_a \dot M_{\rm cross}$
where we relate the two rates with an accretion efficiency $f_a$.

In section \ref{sec:torque_acc} we estimated that accretion 
must remain below a critical accretion rate $\dot M_{cr} \sim 2 \times 10^{10}$ km/yr for Pan,
to maintain tidal lock (otherwise the equatorial ridge would be evenly distributed
about the equator).  Using equation \ref{eqn:Mor} we estimate the critical 
 accretion efficiency $f_{a,cr}$ that gives the critical accretion rate $\dot M_{cr}$ onto the moon, 
\begin{align}
f_{a,cr} &=  \frac{ \dot M_{cr}}{\dot M_{\rm cross}} \nonumber \\
& \sim 3 \times 10^{-5} 
\left( \frac{\dot M_{cr} }{10^{10}\ {\rm kg/yr} } \right)
 \left( \frac{1 \ {\rm km/yr}}{da_{\rm mig}/dt}\right)
\left( \frac{ 400 \ {\rm kg\ m}^{-2}}{\Sigma_{\rm ring}}\right).   \label{eqn:acr}
\end{align}
If the migration rate is only about 1 km/s (which would allow the moon
to stay in the rings longer) then 
the accretion efficiency would be similar to the $f_a \sim 10^{-4}$ we measured for particles accreting
from a ring 
near an eccentric planet per orbit (shown in each panel in Figure \ref{fig:dens}).
In a migrating moon scenario for equatorial ridge accretion, there would be a relation between accretion 
efficiency, the gap edge profiles, the migration rate, the disk density and the impact angle distribution
of accreting material.
Orbital resonances with other moons would also  affect  these quantities.
If migration is not steady then as a consequence we expect variations
in the angular distribution of accreting material.
Such variations might  account for some of the morphology, such
as thickness variations or narrow streaks that are present on Pan's equatorial ridge.


A problem with the migration scenario for equatorial ridge accretion is that for moon
and ring particles in circular orbits, we did not find an explanation for 
the multiple lobes on Pan's equatorial ridge (see section \ref{sec:corb}).  Accretion 
from an asymmetric gap while on an eccentric
orbit is more likely to account  for Pan's ravioli-shaped ridge (explored in
section \ref{sec:c_edge}).  
Eccentricity damping while migrating is expected to be rapid
\citep{rein10,bromley13}.
However, if orbital resonances with
other moons are encountered during migration then perhaps 
migration and eccentricity excitation could be concurrent.
Eccentricity excitation from orbital resonance would require some migration,
and while migration is taking place, 
orbital resonances would be encountered.
So it may not be possible to cleanly separate or differentiate between these two scenarios. 

Up to this point we have neglected Daphnis.  The images of Daphnis shown in Figure \ref{fig:moons} 
suggest that Daphnis is elongated or nearly prolate.  However, the body axis ratios and silhouettes 
by \citet{thomas20} of Daphnis are not significantly different than those of Atlas.  
Daphnis has a second ridge that is at a latitude of about $22^\circ$ N, centered at about $80^\circ$ W and extends about $\sim 90^\circ$  in longitude \citep{buratti19,thomas20}.  
This second ridge could have been accreted while the moon was tilted with
respect to the orbital plane.  
If the moon were 
nearly prolate, it would more easily reorient itself by rotating about its long axis while accreting.
We have explored how the moon can slowly tidally realign by rotating in the equatorial plane
while accreting.  If Daphnis were nearly prolate, perhaps it 
could also roll or vary in obliquity during accretion.   

In summary, we have excluded impact and spin-orbit resonance related scenarios for
accretion of ornamental equatorial ridges. 
Scenarios involving moon migration and eccentricity excitation are promising.
A type III migration-mediated mechanism for equatorial ridge accretion 
has the advantage that material
must cross the gap for migration to take place, the gap hosting
the moon would be narrower, facilitating accretion, and  the gap would probably be 
asymmetric, facilitating accretion of asymmetric equatorial structures.
However, accretion of multiple lobes seems to require moon orbital eccentricity 
which is usually rapidly damped during migration.
Eccentricity excitation due to orbital resonance with another moon
could force a moon into a gap edge,
facilitating accretion; however, we are not sure if  
the gap would remain sufficiently narrow for accretion to continue for the required accretion duration of $\sim 10^5$ years.
To put these moons into
orbital resonance, some migration is needed, and during an epoch
of migration, orbital resonances might be encountered that would increase
orbital inclination and eccentricity and slow the  migration.  These two scenarios
are probably connected.

\section{Summary and Discussion}
\label{sec:sum}

We have estimated a critical accretion rate, $\dot M_{cr}$ that
would be large enough to pull a moon out of a tidally locked state or a spin-orbit
resonance.    If the moon is not tidally locked (or in a spin-orbit resonance) then we would expect
accretion to be evenly distributed around the equator giving a ridge that lacks divots, depressions
or multiple peaks.
As Pan and Atlas's ridges have non-axisymmetric structures, we infer
that the accretion rates of their ridges primarily occurred at rates below $M_{cr} \sim 
10^{10}$ kg/yr.
From the volume in their equatorial ridges, we infer that the duration of 
 accretion was at least $10^5$ years.
 
Integration of test particles on initially circular orbits shows that only particles 
with orbital semi-major axes satisfying $1.7 \lesssim |\Delta a|/R_H \lesssim 2.5 $ 
are likely to impact a moon on a circular orbit upon their first close approach.  
We find that particles can impact the moon's surface at almost
any longitude on the moon's equator. 

We explored a tidal accretion model that takes into account body shape changes
while maintaining tidal alignment.  If the accretion angle is within 
about $45^\circ$ of the sub-planet or anti-planet line, 
the moon can slowly rotate while it accretes 
due to tidal realignment and giving wedge-shaped lobes on the equatorial ridge.   
Integrated orbits show a nearly linear relation
between impact angle and test particle initial orbital semi-major axis.  
A moon accreting from an external gap edge would accrete more
mass at larger western longitudes on its equator and this too can give
a wedge-shaped equatorial lobe, but in this case the slope (or derivative of radial extent with respect to longitude) 
of an accreted wedge is related to the surface density profile of ring material in the gap edge.

For a moon in eccentric orbit, with eccentricity $e>e_H$ or $e \sim e_H$ 
(where $e_H$ is the Hill eccentricity), test particles in initially circular orbits at a single semi-major axis 
 have a wide distribution of possible impact angles. 
Particles initially exterior to the moon's orbit tend to impact the moon on the trailing side. 
Their impact longitude distribution peaks nearer the sub-Saturn point for test particles 
with initial orbital semi-major axis nearer that of the moon, or with lower $|\Delta a|$.
For moon orbital eccentricity $e \sim e_H$, the accretion stream can have two lobes, one impacting
on the trailing side and the other on the leading side.

Multiple lobes, like those on
 Pan's equatorial ridge, might be accreted from an asymmetric
gap and at orbital eccentricity $e \sim e_H$. 
A depression on the moon's leading side, such as present on Atlas's ridge, 
could be due to accretion from an asymmetric gap, either at low eccentricity
and with a wider gap exterior to the moon's orbit, or at $e > e_H$ and
with a wider gap interior to the moon's orbit.

We exclude impact and spin-orbit resonance related scenarios for
accretion of ornamental equatorial ridges. 
A type III moon migration-mediated mechanism for equatorial ridge accretion 
has the advantage that material
must cross the gap for migration to take place, the gap hosting
the moon would be narrower, facilitating accretion, and the gap would be 
asymmetric, facilitating accretion of asymmetric equatorial structures.
Eccentricity excitation due to orbital resonance with another moon
could push a moon into a gap edge,
facilitating accretion, however   
the gap may widen afterward, cutting off accretion. 
A scenario for  accretion while in orbital resonance has the advantage
that the orbital eccentricity might be maintained for the required $10^5$ year duration, 
otherwise eccentricity would be damped rapidly by driving spiral density waves \citep{hahn08,rein10,bromley13}.
To enter orbital resonance with another moon, some migration is needed, and during an epoch 
of migration, orbital resonances would be encountered and possibly maintained. These two scenarios
are probably connected. Migration, eccentricity and inclination excitation could be concurrent.

In this study we have considered orbits near point masses, and neglected collisions
between accreting particles, scattering upon impact, and mass redistribution on the moon's 
surface or in the accreted ridge interior. 
We assumed that the accretion stream originates from material in nearly circular orbits, 
which would be a reasonable approximation in a quiescent high opacity ring.
Future studies could study the role of the complex local gravity field and collisions between
particles that could
occur in the accretion stream.  We primarily looked at orbits in the equatorial plane,
but the inclination distribution of ring particles affects the 
thickness of accreted equatorial ridges \citep{charnoz07} and could also affect the migration
rate \citep{bromley13}.    As test particles were begun in circular orbits
we have neglected the role of waves in ring edges \citep{cuzzi85,weiss09} that might be correlated
with the phase of the moon in its orbit.  These too would affect the angular distribution
of impacts on a moon's surface.  We have neglected rotational tilt and libration
of the moon that might be excited during accretion.
Because impact craters are visible on Pan and Atlas's equatorial ridges  \citep{buratti19}, we have assumed that accretion was higher in the past.  
However, the Encke gap is not empty \citep{hedman13}, so accretion could be ongoing, but at a low rate. 

The N-body simulations of equatorial ridge accretion by \citet{charnoz07} lacked collisions between ring particles and were run only for about $10^4$ orbits.  As accreting ring particles were massless, migration
did not take place. 
These numerical issues affect the gap density profile and the 
orbital eccentricity and inclination of distributions of particles in the gap edges, and 
as a consequence, the latitude and longitude
distribution of the simulated impacts on the moon.
The N-body simulations by \citet{bromley13} showed type III migration but did not track the near moon environment to follow equatorial ridge growth.  Neither set of simulations
included additional moons that could excite orbital resonances. It would be interesting to carry out more complex and sophisticated simulations that might test some of our speculations on equatorial ridge growth.



\begin{table*}
\ifpreprint
\else
\small
\fi
\begin{center}
\caption{\large Moon measured properties \label{tab:moon}}
\begin{tabular}{@{}lllllll}
\hline
Quantity    & symbol & units  & Pan   & Daphnis & Atlas \\
\hline
Orb. semi-major axis & $a$& $R_S$ & 2.22  & 2.26    & 2.29  \\
Orb. Eccentricity         & $e$&       & $1.44 \pm 0.54 \times 10^{-5}$ & 
$3.31 \pm 0.62 \times 10^{-5}$& 0.012 \\
Orb. Inclination          & $i$ &   degrees    
& $1 \pm 4 \times 10^{-4}$      
& $3.6\pm 1.3 \times 10^{-3}$  & $3.1 \times 10^{-3} $ \\
Orb. Rotation period & $P$   & days    & 0.575 & 0.594   & 0.602 \\
Mass            & $M$&$10^{16}$\ kg & 0.495 & 0.0077  & 0.66  \\
Density         & $\rho$&kg\ m$^{-3}$& $400\pm 31$ & $276\pm 144$     & $412 \pm 19$  \\
Mean body radius     & $R_m$ &  km     & $13.7\pm 0.3$  & $3.9\pm 0.5$     & $14.9\pm 0.2$  \\
Body semi-major axis&$a_{\rm body}$&km     & $17.3\pm 0.2$  & $4.9\pm 0.3$     & $20.4\pm 0.1$  \\
Body semi axis      &$b_{\rm body}$&km     & $14.1\pm 0.2$  & $4.2\pm 0.8$     & $17.7\pm 0.2$  \\
Body semi-minor axis&$c_{\rm body}$&km     & $10.5\pm 0.5$  & $2.8 \pm 0.6$     &  $9.3\pm 0.3$  \\
Volume fraction  &  $f_v$ & \% & 10 & 1 & 25 \\
Core axis ratio  & $c_{\rm core}/a_{\rm core}$ &  & 0.65 & & 0.54 \\
Core axis ratio  & $c_{\rm core}/b_{\rm core}$ &  & 0.78 & & 0.58 \\
Core axis ratio  & $b_{\rm core}/a_{\rm core}$ &  & 0.83 & & 0.93 \\
\hline
\end{tabular}
\end{center}

{Notes: Here $R_S = 60268 $ km is the equatorial radius of Saturn.  The mass for Atlas is taken from \citet{jacobson08} and those for Pan and Daphnis from \citet{porco07}.   Orbital elements are by \citet{jacobson08}.
Mean density $\rho$,  
mean radius $R_m$,  body semi axes $a_{\rm body},b_{\rm body},c_{\rm body}$ 
volumes of the equatorial ridges divided by total moon volume, $f_v$, are  by \citet{thomas20}. 
Core axis ratios are estimates of the body axis ratios after removing the equatorial ridge and also by  \citet{thomas20}.  We omit these for Daphnis as the volume fraction in its ridges is small.
 }
\end{table*}

\begin{table*}
\begin{center}
\caption{\large  Moon computed properties \label{tab:computed}}
\begin{tabular}{@{}lllllll}
\hline
\hline
              & symbol & units  & Pan   & Daphnis & Atlas \\
              \hline
Hill radius   &  $R_H$ & km     & 19.09 & 4.85 & 21.02 \\
Hill velocity & $v_H$   & m/s    &  2.4    &  0.6  &  2.5     \\
Hill eccentricity & $e_H$ &   & $1.4\times 10^{-4}$ & $3.6\times 10^{-5}$ & $1.5 \times 10^{-4}$ \\
Ratio of fill &$a_{\rm body}/R_H$ &      & 0.91  & 1.01 & 0.97 \\
Time to spin down & $t_{\rm despin}$ & yr & $9 \times 10^4 $ & $1 \times 10^6 $& $9 \times 10^4 $ \\
Time to spin down & $t_{\rm despin}/P$ &  & $6 \times 10^7 $& $9 \times 10^8$ & $6 \times 10^7 $\\
Accretion to tidal torque ratio   &$\frac{T_{\rm acc}}{T_2}\frac{Mn}{\dot M} $   & & $10^9$ & $10^{10}$ & $10^9$ \\
Critical accretion rates  & $\dot M_{cr}$ & kg/yr & $2 \times 10^{10} $ & $2 \times 10^7$ &  $3 \times 10^{10} $ & \\
Critical accretion time scale & $M f_v/\dot M_{cr}$ & yr & $2 \times 10^4$ & $4\times 10^4$ & $6 \times 10^4$ \\
\hline
\hline
\end{tabular}
\end{center}
{Notes: The Hill radius $R_H = a(M/(3M_*))^\frac{1}{3}$ is computed from values in Table \ref{tab:moon}.  
The Hill velocity $v_H = a R_H$. 
The Hill eccentricity $e_H = R_H/a$.  
The ratio of fill is the body semi-major axis divided by the Hill radius, $a_{\rm body}/R_H$.
Here $P$ is the orbital period.  Spin-down times are computed
using shear modulus times dissipation factor $\mu_{\rm shear} Q \sim 10^{11}$ Pa
and equation \ref{eqn:t_despin}.
The ratio of accretion to tidal torque $\frac{T_{\rm acc}}{T_2}\frac{Mn}{\dot M}$ is computed using equation \ref{eqn:Tacc_ratio}. 
The critical accretion rate, $\dot M_{cr}$,  that can pull the moon out of the spin synchronous state
is estimated using equation \ref{eqn:Mcrit}.
}
\end{table*}

\vskip 1 truein 
Acknowledgements.

This material is based upon work supported in part by NASA grant 80NSSC17K0771.
We are grateful to Randall C. Nelson, Arnav Sharma, John Siu, Alex Gunn, Matt Tiscerano and Aurelien Crida for discussions and suggestions that inspired and significantly improved our manuscript.
We thank Matt Tiscerano for providing us with information about the orientation of Cassini images.
This work used the PDS Ring-Moon Systems Node's OPUS search service. 

\vskip 2 truein

{\bf Bibliography}

\bibliographystyle{elsarticle-harv}
\bibliography{refs_tides}

\appendix
\section{Changes to body shape and tidal alignment}
\label{ap:mom}

We describe how we find the body principal axes in our two-dimensional accretion 
model.  Tidal alignment is mimicked by tilting the body so that
the major axis remains oriented vertically.  
We start with a coordinate frame with $x$ along the long principal
body axis and $z$ along the short principal body axis.
The moment of inertia matrix is diagonal
\begin{equation}
{\bf I} =  \begin{pmatrix}
A & 0  & 0\\
0 & B & 0 \\
0 & 0 & C  \end{pmatrix} 
\end{equation}
with $A< B< C$.

We now consider the same body that is rotated counter-clockwise by angle $\phi$ in the $xy$ plane.
The moment of inertia matrix is no longer diagonal.
It is convenient to define a 2-dimensional matrix 
\begin{equation}
\tilde {\bf I} = \begin{pmatrix}  I_{xx} & I_{xy} \\ 
 			I_{xy} & I_{yy} \end{pmatrix} .
\end{equation}
This same body now has 
 $xy$ components of the moment of inertia matrix 
\begin{align}
\tilde {\bf I} 
&= 
 \begin{pmatrix}
	\cos \phi & -\sin \phi  \\
	\sin \phi  &  \cos \phi 
	\end{pmatrix}
\begin{pmatrix}	
	A & 0 \\
	0 & B \end{pmatrix}
\begin{pmatrix}
	\cos \phi  &  \sin \phi  \\
 	-\sin \phi  &  \cos \phi  
	\end{pmatrix}\nonumber \\
	 &= 
\frac{A+B}{2}  \begin{pmatrix} 
	1 & 0 \\ 0 & 1
 		\end{pmatrix}  + 
\frac{A-B}{2}  \begin{pmatrix} 
 	\cos(2 \phi) & \sin (2 \phi) \\
	\sin(2 \phi) & -\cos (2 \phi) 
 		\end{pmatrix} 	 .
\end{align}
The $I_{zz}$ component of the moment of inertia matrix is unchanged by rotation
in the $xy$ plane.
This relation for $\tilde {\bf I}$ can be inverted to 
 compute the rotation angle $\phi$ from the components
of the moment of inertia matrix,
\begin{equation}
\phi = \frac{1}{2} {\rm arctan2}\left(2I_{xy}, I_{xx}  - I_{yy}  \right) .  \label{eqn:phi}
\end{equation}
This rotation angle determines the orientation of a principal body axis in the $xy$ plane.
The unit vector $(\cos \phi, \sin \phi)$ is 
 an eigenvector of $\tilde {\bf I}$ but it could have   
 eigenvalue $A$ or $B$.
To determine the orientation of the longest body axis, we compute the two eigenvalues of $\tilde {\bf I}$ 
using its eigenvectors 
\begin{align} 
\lambda_1 &= \left| \tilde {\bf I} \begin{pmatrix} \cos \phi \\ \sin \phi \end{pmatrix} \right|  \qquad 
\lambda_2 &= \left| \tilde {\bf I} \begin{pmatrix} \cos (\phi+\pi/2) \\ \sin (\phi+\pi/2) \end{pmatrix}\right|.
\end{align}
The angle associated with the smaller eigenvalue gives the orientation angle of
the long body axis.  We use this computation to determine the orientation of
a tidally aligned body. 

Accretion will cause the moment of inertia matrix to vary. 
With accretion at a fixed angle $\theta_a$ will the body orientation $\phi$ increase
or decrease?
We take the derivative of the rotation angle  $\phi$ with respect to changes in the components 
of the moment of inertia matrix
\begin{align}
\delta \phi  = 
\frac{
-    I_{xy} (\delta I_{xx} - \delta I_{yy})  + (I_{xx} - I_{yy}) \delta  I_{xy}    }
{4 I_{xy}^2 + (I_{xx} - I_{yy})^2} .
\end{align}
We assume that the body is tidally aligned so 
 $\tilde {\bf I}$ is originally diagonal.  Then $I_{xy}$ vanishes, and the above equation becomes
\begin{align}
\delta \phi =
   \frac{ \delta  I_{xy}}{I_{xx} - I_{yy}} .  \label{eqn:deltaphi}
\end{align}
We assume that $I_{yy} < I_{xx}$ for tidal alignment of the long
axis vertically. 

We add mass at radius $r_a$ and at accretion angle $\theta_a$ from vertical.
The mass is added  at position $(x,y) = r_a(-\sin \theta_a, \cos \theta_a)$, 
 giving a change in the xy components of the moment of inertia matrix 
\begin{equation}
\delta \tilde {\bf I} =  \frac{dm}{2} r_a^2 
\begin{pmatrix}
	1 + \cos (2 \theta_a) & \sin (2\theta_a)  \\
	 \sin (2\theta_a)   &  1- \cos(2 \theta_a )
\end{pmatrix}.
\end{equation}
This gives a change to the orientation angle of the principal body axis 
\begin{equation}
\delta \phi =  \frac{dm}{2} r_a^2 \frac{ \sin (2\theta_a) }{ (I_{xx} - I_{yy}) }.
\end{equation}
Since the angle adjustment depends on the difference in the diagonal components of
the moment of inertia matrix 
it is useful to compute the change in  $I_{xx} - I_{yy}$  caused by accreting the mass $dm$,
\begin{align}
\delta (I_{xx} -  I_{yy}) = dm\ r_a^2 \cos (2 \theta_a).  \label{eqn:deltaIxxyy}
\end{align}
If $|\theta_a| < \pi/4$ then the difference in the eigenvalues of the moment
of inertial matrix is enhanced and the body becomes more elongated.
This in turn affects the size of $\delta \phi$.  As mass is added,
the size of the angular shift in body orientation decreases.  A wedge-shaped lobe
is accreted.
If $|\theta_a| > \pi/4$ then the body becomes rounder and the size
of the angular shift increases as mass accretes.  Mass is accreted at all longitudes. 
These trends are for a fixed accretion radius $r_a$ and here we have
ignored its dependence on angle.  For more elongated bodies, the angular
shift would be smaller than estimated with an angular independent accretion
radius $r_a$.

\end{document}